\documentclass[aps,pre,twocolumn,amsmath,amssymb]{revtex4}
\usepackage{graphicx}

\newcommand{\var}{\mbox{Var}}
\newcommand{\xb}{\mbox{\bf x}}
\newcommand{\rb}{\mbox{\bf r}}
\newcommand{\ub}{\mbox{\bf u}}
\newcommand{\Ub}{\mbox{\bf U}}

\newcommand{\Rb}{\mbox{\bf R}}

\newcommand{\Db}{\mbox{\bf D}}
\newcommand{\Ab}{\mbox{\bf A}}
\newcommand{\Pc}{{\cal P}}
\newcommand{\del}{\mbox{\boldmath{$\nabla$}}}
\newcommand{\Pe}{\mbox{P\hspace{-1pt}e\hspace{1pt}}}

\begin{document}



\title{ A Theory of Cooperative Diffusion in Dense Granular Flows }



\author{Martin Z. Bazant}
\email{bazant@math.mit.edu}
\affiliation{
Department of Mathematics, Massachusetts Institute of Technology,
Cambridge 02139 }%

%

\date{May 8, 2004}

\begin{abstract}

Dilute granular flows are routinely described by collisional kinetic
theory, but dense flows require a fundamentally different approach,
due to long-lasting, many-body contacts. In the case of silo drainage,
many continuum models have been developed for the mean flow, but no
realistic statistical theory is available.  Here, we propose that
particles undergo cooperative displacements in response to diffusing
``spots'' of free volume. The typical spot size is several particle
diameters, so cages of nearest neighbors tend to remain intact over
large distances.  The spot hypothesis relates diffusion and
cage-breaking to volume fluctuations and spatial velocity
correlations, in agreement with new experimental data. It also
predicts density waves caused by weak spot interactions. Spots enable
fast, multiscale simulations of dense flows, in which a small,
internal relaxation enforces packing constraints during spot-induced
motion. In the continuum limit of the model, tracer diffusion is
described by a new stochastic differential equation, where the drift
velocity and diffusion tensor are coupled non-locally to the spot
density.  The same mathematical formalism may also find applications
to glassy relaxation, as a compelling alternative to void (or hole)
random walks.

\end{abstract}

\maketitle

\section{Introduction}

In spite of a venerable engineering
literature~\cite{schofield,brown,nedderman}, the study of granular
materials is attracting a growing community of
physicists~\cite{jaeger92,mehta94,jaeger96,degennes99,halsey02}, who
sense that basic principles remain to be discovered~\cite{edwards01},
including ``the chance to reinvent statistical mechanics in a new
context''~\cite{kadanoff99}. Of course, individual grains obey the
laws of classical physics, but in many ways their collective behavior
defies standard assumptions in statistical thermodynamics and
hydrodynamics. Static properties of random packings, such as
geometrical correlation functions~\cite{torquato,torquato02}, the
jamming
transition~\cite{liu98,torquato00,torquato01,kansal02,ohern02,ohern03},
and local force
distributions~\cite{bouchaud95,coppersmith96,edwards98a,mueth98,blair01,landry03,ngan03}
are far from fully understood, but granular flows may pose even more
fundamental open questions.

Naturally, the most attention has focused on the regime of fast,
dilute flow because it is closest to the familiar molecular-fluid
state. In this regime, grains to undergo simple random walks due to
binary (or possibly many-body) collisions, which differ from those in
normal fluids only by being inelastic. From these postulates,
continuum equations for hydrodynamics and heat transfer can be
formally derived from various modifications of Boltzmann's kinetic
theory of
gases~\cite{savage79,savage81,jenkins83,haff83,lun84,prakash91,hsiau93b,dufty01}.
Due to inelastic collisions, external forcing must supply enough
kinetic energy to approximately satisfy the assumption of thermal
equilibrium. The resulting definition of a time-dependent, local
``granular temperature'' in terms of velocity fluctuations is somewhat
controversial, since classical statistical mechanics may break down
with inelastic collisions~\cite{kadanoff99}. For example,
non-extensive Tsallis statistics~\cite{tsallis88} can be derived from
the simple ``inelastic Maxwell model'' of a granular
gas~\cite{sattin03}.  In any case, it is clear that kinetic theory
breaks down with increasing density~\cite{natarajan95,choi04}, when
long-lasting many-body contacts are formed. Due to gravity, such is
usually the case in granular materials.

In seeking new and relevant physics, therefore, it seems more fruitful
to focus on the opposite regime of slow, dense flow --- closer to the
``granular solid'' than the ``granular
gas''~\cite{jaeger96}. Mean flow fields in draining silos, incline
chutes, and shear cells have been studied extensively, but a simple
unified description remains elusive. Classical continuum models of
dense flows are based on plasticity theory from soil
mechanics~\cite{schofield,nedderman}. Although they remain popular in
engineering, these models predict complicated patterns of velocity
discontinuities (``rupture surfaces''), consistent with some
experiments~\cite{pariseau70,blair73,michalowski84} but not
others~\cite{nedderman79,tuzun79,medina98a,samadani99,choi04}, and can
lead to violent instabilities~\cite{schaeffer87,pitman87}.  On the
other hand, another class of early models for silo drainage (discussed
below in section) involves only simple geometrical
considerations~\cite{lit58,lit63a,lit63b,mullins72,mullins74,nedderman79}.
Recent attempts to develop continuum models for dense flows have
introduced a remarkable variety of physical postulates, such as a
temperature-dependent viscosity~\cite{savage98}, density-dependent
viscosity~\cite{bocquet02,losert00}, non-local stress propagation
along hypothetical arches~\cite{mills99}, self-activated shear events
due to non-local stress fluctuations~\cite{pouliquen96,pouliquen01},
coexisting liquid and solid micro-phases governed by a Landau-like
order parameter~\cite{aranson99,aranson01}, and shear-induced local
rearrangements mediated by free-volume
kinetics~\cite{lemaitre02,lemaitre02c}.

Putting aside the question of which of these continuum models best
describes the mean velocity, we focus here on the stochastic motion of
individual particles outside the collisional regime, for which no
theory is available.  Even in experiments, little attention has been
paid to this fundamental issue, perhaps due to the practical
difficulty of tracking individual particles in dense flows.  Due to
recent advances in digital video imaging, however, thousands of
particles near the wall of a dense granular flow can now be tracked
simultaneously with very high
resolution~\cite{horluck99,horluck01,choi04} (well beyond previous
standards~\cite{natarajan95,medina98a,medina98b}). Combining these
techniques with magnetic resonance imaging and x-ray tomography can
also yield three-dimensional information~\cite{mueth00}. Complete
microscopic motion in the bulk can also be obtained from granular
dynamics simulations, now involving up to hundreds of thousands of
particles in three
dimensions~\cite{makse99,makse02,hirshfeld01,silbert01,silbert02b,landry03}.
These revolutionary capabilities should aid in developing a 
statistical theory of dense granular flows.

In some sense, work along these lines has already begun with attempts
to define a meaningful temperature for dense granular systems. The
notion of ``Edwards
temperature''~\cite{edwards91,edwards94,edwards98}, analogous to
structural temperatures in glasses~\cite{kurchan01}, has attracted
increasing attention over the past decade.  The Edwards temperature
differs from its Boltzmann counterpart in that the entropy is based on
static random packings at a given volume, rather than thermal
configurations at a given energy. As such, it may be related to the
compaction of powders under vibration~\cite{edwards98,nowak98}.  Some
support for the idea has come from recent molecular dynamics
simulations verifying an effective Einstein relation between mobility
and diffusivity in an artificial shear
flow~\cite{makse02}. Nevertheless, it is not clear how the Edwards
equilibrium ensemble relates to hydrodynamics and diffusion (if at
all) since no analog of the Boltzmann equation of kinetic theory has
been proposed.  This would seem to require replacing the usual
hypothesis of a random walk, due to ballistic motion between
instantaneous collisions, with a new dynamical mechanism.

Thus we arrive at the central question of this work:
\begin{quote}
{\it Is there a simple
analog of the random walk for a dense granular flow? }
\end{quote}
Roughly a century after random-walk theory was first developed, the
original idea of a single entity, such as a
mosquito~\cite{pearson1905}, sound-wave
amplitude~\cite{rayleigh1880,rayleigh1905}, Brownian
particle~\cite{einstein1905,smoluchowski1906}, or stock
price~\cite{bachelier1900}, undergoing random displacements remains in
most subsequent generalizations~\cite{wax,hughes,bouchaud}. For
example, the ``continuous-time random walk'' of Montroll and
Weiss~\cite{montroll65,weiss83} and related models~\cite{hughes},
which describe anomalous diffusion in semi-conductors~\cite{scher75},
turbulent flows~\cite{schlesinger87} and many other
systems~\cite{metzler00} via a random waiting time between steps, also
assume an independent random walker, unaffected by its neighbors. The
same is true of the ``persistent random walk'' of
F\"urth~\cite{furth1920} and Taylor~\cite{taylor1921} which introduces
auto-correlations between steps, e.g. to model a transition from
short-time ballistic to long-time diffusive
motion~\cite{corrsin74,weiss02}.

Of course, there are correlations between different random walkers in
any condensed-matter system (or financial market~\cite{bouchaud}), but
in most cases it is reasonable to assume that the diffusion
of a single particle is independent of all the others.  This universal
independence of single-particle trajectories is often related to 
attaining the thermodynamic limit, in which an enormous number ($N
\approx 10^{23}$) of particles undergo very frequent collisions
($\omega = 10^{14}$ Hz) at finite temperature (in the classical
sense). As a result, short-range many-body interactions have little
affect on tracer diffusion at macroscopic length and time scales,
aside from influencing parameters, such as the diffusivity.

In stark contrast, the ``cages'' of neighboring particles in a dense
granular material may remain at least partially intact for the entire
duration of slow flow~\cite{choi04}, due to strong internal energy
dissipation and much smaller system sizes (e.g. $N\approx 10^5$). Even
at macroscopic length and time scales, therefore, particles must
diffuse together in a cooperative fashion, and it is not clear what
kind of statistics might describe their collective motion.
Boltzmann's kinetic theory of gases does not appear to be the
appropriate starting point.

\begin{figure*}
\begin{center}
\mbox{
\includegraphics[width=5.5in]{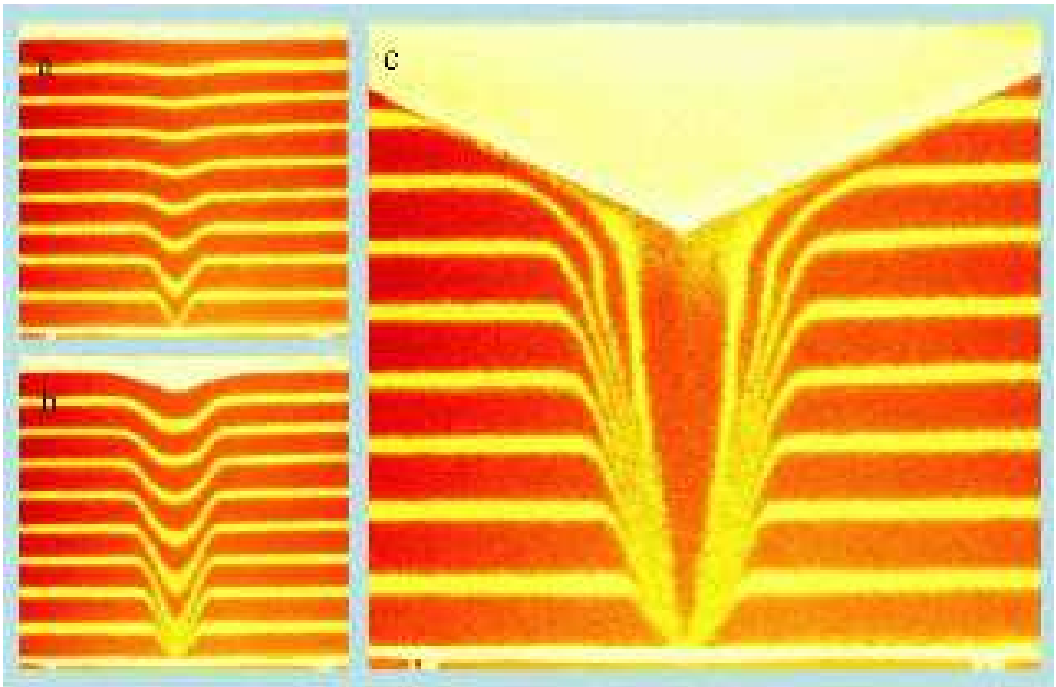}
}
\caption{ \label{fig:azadeh} Sequence of experimental photographs of a
$d=3$mm glass beads draining from a quasi-two-dimensional silo
(courtesy of A. Samadani and A. Kudrolli~\cite{samadani99}). Particles
are colored according to their initial positions in equally spaced
horizontal layers when the flow began.  }
\end{center}
\end{figure*}

In this article, we propose a simple mathematical model for
cooperative diffusion and apply it to slow granular drainage from a
silo, as shown in Fig.~\ref{fig:azadeh}. After summarizing our results
in section~\ref{sec:summary}, we begin by showing that the only
existing model for granular diffusion, based on the concept of
diffusing voids, is fundamentally flawed and draw clues from
experiments as to what could be missing. In section~\ref{sec:spot}, we
introduce the general concept of a ``spot'' of cooperative diffusion,
and in section~\ref{sec:expt} we show that it is suffices to predict a
wide variety of experimental data.  In section~\ref{sec:sim}, we
briefly discuss Monte Carlo simulations with the model and spot-based
multiscale algorithms. In section~\ref{sec:math}, we derive general
continuum equations for tracer diffusion in the spot model, focusing
in section~\ref{sec:spotcont} on the case of granular drainage.  We
conclude in sections~\ref{sec:outlook}--\ref{sec:concl} by discussing
how the theory might have broad applicability to cooperative diffusion
in amorphous materials.

\section{ Summary of Results }
\label{sec:summary}

As a guide for the reader, we list our main results:
\begin{enumerate}
\item A new microscopic picture of dense granular flows is presented
in Figs.~\ref{fig:rising_spots} and \ref{fig:spot}.  Rising spots
of free volume cause strongly correlated random displacements of
blocks of neighboring particles. Each random spot path corresponds to
a thick chain of reptating particles.
\item In section~\ref{sec:spot}, simple calculations show
how the spot hypothesis resolves the ``paradox of granular diffusion''
illustrated in Fig.~\ref{fig:azadeh}: Particles diffuse
several orders of magnitude more slowly than free volume.
\item In section~\ref{sec:expt},  predictions
of the P\'eclet number, diffusion length, and cage-breaking distance,
consistent with experimental data, are made by assuming volume
fluctuations of order one percent.  The vertical diffusion length of
free volume, which is difficult to measure experimentally, is also
predicted.
\item Direct experimental evidence for spots, through spatial
velocity correlations, is shown in Fig.~\ref{fig:Corr}.
\item The necessity of weak repulsive interactions between spots
naturally explains experimental density waves, sketched in
Fig.~\ref{fig:waves}, which propagate upward in wide funnels and
possibly downward in narrow funnels.
\item Monte Carlo simulations with the simplest version of the model,
shown in Figs.~\ref{fig:spot_layers} and \ref{fig:spot_interface},
accurately describe tracer-particle dynamics and some aspects of
many-body correlations, although unphysical density fluctuations grow
in regions of high shear.
\item The density may be stabilized by minor relaxations within spots
to enforce packing constraints, which amounts to a very efficient
multiscale simulation algorithm for granular flow. These extra
fluctuations, sketched in Fig.~\ref{fig:spot3}, may be responsible for
the nontrivial (sub-ballistic) super-diffusion seen in experiments at
scales less than a particle diameter. 
\item In section ~\ref{sec:math}, the mean-field continuum limit of
the model is analyzed, starting from non-local stochastic differential
equation, Eq.~ (\ref{eq:tracer}). The result is a
Fokker-Planck equation for tracer diffusion, Eq.~(\ref{eq:fp}), whose
coefficients in Eqs.(\ref{eq:up}) and (\ref{eq:Dp}) are coupled
non-locally to the spot (or free-volume) density.
\item A general tensor relation between transport coefficients,
  Eqs.~(\ref{eq:Cp}), is derived, which relates spatial velocity
  correlations and the diffusivities of particles and free volume.
\end{enumerate}

\section{ Towards a Statistical Theory }
\label{sec:old}

\subsection{ The Kinematic Model for the Mean Velocity }

In the 1970s, engineers developed a very simple ``Kinematic Model''
for the mean velocity field in a steadily draining
silo~\cite{nedderman}, such as the flow illustrated in
Fig.~\ref{fig:azadeh}. It is based on the following constitutive
law, suggested by Nedderman and Tuzun~\cite{nedderman79},
\begin{equation}
\ub = b\, \del_\perp v  \label{eq:law}
\end{equation}
which postulates that the horizontal velocity, $\ub$, is proportional
to the horizontal gradient, $\del_\perp$, of the downward velocity
component, $v$ (i.e. the shear rate). We refer to the constant of
proportionality, $b$, as the ``diffusion length" of free volume for
reasons soon to become clear. The intuition behind this relation is
that particles should tend to drift horizontally toward regions of
faster downward flow, which are less dense and more accommodating of
newcomers.  Assuming that these density fluctuations are small enough
to justify steady-state incompressibility yields an equation for the
downward velocity,
\begin{equation}
\frac{\partial v}{\partial z} = b\, \del_\perp^2 v ,  \label{eq:kin}
\end{equation}
which is simply the diffusion equation, where the vertical coordinate,
$z$, plays the role of time.  As such, ``initial conditions'' must be
specified for the downward velocity at the bottom of the silo, and
information propagates upward.

For example, a point source of velocity (a narrow orifice) at $z=0$ in
a quasi-two dimensional silo produces the basic solution,
\begin{equation}
v(x,z) = \frac{e^{-x^2/4 b z}}{\sqrt{4\pi b z}}   \label{eq:meanflow}
\end{equation}
for the case of constant $b$. The flow is effectively confined to a
region of parabolic streamlines due to the diffusive scaling, $\Delta
x \propto \sqrt{z}$. As show in Fig.~\ref{fig:azadeh}, this prediction
is in rather good agreement with experimental measurements of the bulk
flow (well below the free surface) for large, dry
grains~\cite{mullins74,tuzun79,medina98a,samadani99,choi04}.  (For
fine powders and soils, the model breaks down, presumably due to
cohesion between grains and/or the influence of the interstitial
fluid~\cite{nedderman,tuzun79}.)

The fact that a single fitting parameter, $b$, suffices to reproduce
the entire flow field reasonably well should be viewed as a major
success of the Kinematic Model. Nevertheless, the model fell from
favor in the 1980s, in part because the diffusion length seems hard to
predict {\it a priori}. Although $b$ is consistently on the order of a
several particle diameters, its precise value depends somewhat on the
particle size and flow symmetry~\cite{nedderman}. (It also varies with
the humidity in cases outside the pure "granular"
regime~\cite{samadani99}.) A Gaussian fit to the downward velocity
profile also seems to require a larger value of $b$ in the upper
region of the silo, where the flow is more plug-like. This was also
seen as problematic, although the model does not really require that
$b$ be a constant.

Perhaps a deeper reason for the dissatisfaction with the Kinematic
Model is the natural prejudice against a theory seemingly devoid of
``physics". The model depends only upon geometry, via the diffusion
length and the silo shape, and not on any of the usual physical
variables, such as momentum, mass, energy, temperature, stress,
etc. In contrast, more complicated models from soil mechanics involve
a stress-based yield criterion, which generally leads to bulk
discontinuities in stress and velocity (``rupture
zones'')~\cite{nedderman,prakash91}. It is not clear that these models
provide a better overall description of silo drainage, but in any case
it seems worth taking a fresh look at the simpler Kinematic Model and
its possible microscopic justification.

\subsection{ The Void Model for the Mean Velocity }

\begin{figure}
\begin{center}
\mbox{ \includegraphics[width=1.5in]{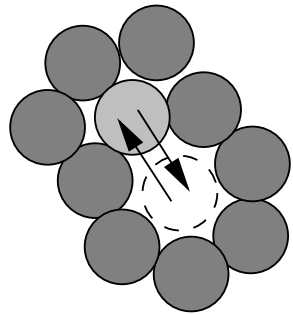} } \caption{
\label{fig:void} The Void Hypothesis I. A single
particle drops independently from one available cage to
another~\protect\cite{lit58,lit63a,lit63b} by (equivalently)
exchanging with a rising void
~\protect\cite{mullins72,mullins74,hong91}. As a result, cage breaking
occurs at the scale of the particle diameter. }
\end{center}
\end{figure}

Although Nedderman and Tuzun introduced the macroscopic view of
Equation~(\ref{eq:law}) as a constitutive relation and
Eq.~(\ref{eq:kin}) as conservation law~\cite{nedderman79}, both
equations had been derived years earlier by
Litswiniszyn~\cite{lit58,lit63a,lit63b} and
Mullins~\cite{mullins72,mullins74} from a statistical model which
remains, to this day, the only microscopic description of silo
drainage. The basic hypothesis is sketched in Figure~\ref{fig:void}.
Litwiniszyn first suggested in 1958 that individual hard-sphere
particles diffuse downward through a fixed array of available
``cages'', which yields the Kinematic Model for the mean velocity in
the continuum limit~\cite{lit58,lit63a,lit63b}.

\begin{figure}
\begin{center}
\mbox{ \includegraphics[width=2.7in]{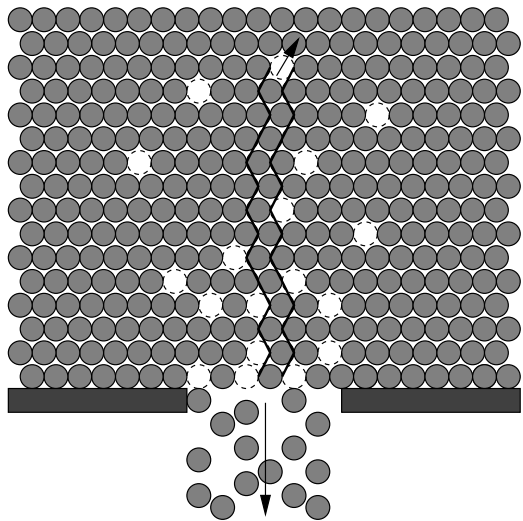} }
\caption{ \label{fig:rising_voids} The Void Hypothesis II. On a
lattice~\protect\cite{hong91}, a void trajectory emanating from the
orifice causes a reptating chain of downward particle displacements
(thick lines), each at the scale of the particle diameter.  }
\end{center}
\end{figure}

Independently, Mullins in 1972 derived the Kinetic Model from the
hypothesis that that particles move passively downward in response to
the upward diffusion of ``voids'' emanating from the silo
orifice~\cite{mullins72,mullins74}. Although formally equivalent to
Litswiniszyn's particle-based model~\cite{mullins79}, Mullins'
void-based model introduces a fundamentally different perspective,
analogous to vacancy diffusion in crystalline solids, which we will
find quite suggestive. (Note that the same mechanism of ``hole
diffusion'' was first proposed in the free-volume theory of molecular
liquids and glasses~\cite{eyring36,cohen59}, discussed below in
section~\ref{sec:glassy}.)

The Kinematic Model is easily derived from the continuum limit of the
Void Model, assuming that voids do not interact.  According to the
Central Limit Theorem~\cite{hughes}, the void concentration, $\rho_v$,
(or probability density) tends toward a Gaussian profile satisfying
the diffusion equation,
\begin{equation}
\frac{\partial \rho_v}{\partial z} = b \, \del^2_\perp \rho_v ,  \label{eq:rhov}
\end{equation}
where the parameter, $b$, is well-defined microscopic quantity, the
void diffusion length (usually assumed to be constant). The mean
particle velocity,
\begin{equation}
\rho_p\, (\ub, -v) = - v_0 \, (-b \,\del_\perp \rho_v, \rho_v) ,
\label{eq:voidflux} 
\end{equation}
is obtained by equating the particle flux density (on the left) with
minus the void flux density (on the right), where $v_0$ is the mean
upward drift velocity of the voids (related to the flow rate), and
$\rho_p$ is the particle density, which is roughly constant in a dense
flow. Equations~(\ref{eq:law}) and (\ref{eq:kin}) follow from
Equations~(\ref{eq:rhov}) and (\ref{eq:voidflux}).

When physicists, Hong and Caram, revisited the Void Model in 1991, 
they performed computer simulations of voids undergoing directed
random walks on a lattice~\cite{hong91}, as illustrated in
Fig.~\ref{fig:rising_voids}. Although they did not perform any mathematical
analysis, the void diffusion length is easily calculated for
any regular lattice:
\begin{equation}
b = \frac{ \var( \Delta \xb_v )}{2 d_h \Delta z_v}  \label{eq:blattice}
\end{equation}
where $\Delta \xb_v$ is the random horizontal displacement (in $d_h=2$
horizontal dimensions) when a void moves up by $\Delta z_v$. For
example, assuming isotropic transition probabilities to
nearest-neighbor sites in the next lattice plane upward, we find $b =
d/4\sqrt{6}$ for the face-centered cubic and hexagon close-packed
lattices and $b = d/4\sqrt{2}$ for the body-centered cubic
lattice. More generally, for any conceivable lattice approximating the
random close packing of hard spheres, the Void Model predicts $b \ll
d$, in contrast to experimental measurements which always yield $b >
d$, e.g. $b = 1.3\, d$~\cite{choi04}, $b = 2.3\, d$ ~\cite{nedderman79},
$b = 3.5\, d$ ~\cite{samadani99}, and $b = 2\, d$-- $4\, d$
~\cite{medina98a}.  This suggests that the microscopic picture of the
Void Model is somehow flawed, even though the mean velocity profile is
quite reasonable.

\subsection{ Particle Dynamics in the Void Model }

More serious problems with the Void Model become apparent when one
considers diffusion and mixing, which it seems has not previously been
done. Mullins and Litwiniszyn were content to use the Void Model as
simply a means to derive the continuum equation ~(\ref{eq:kin}).
Likewise, in spite of doing discrete simulations, Hong and Caram only
measured the mean velocity profile.

\begin{figure}
\begin{center}
\includegraphics[width=\linewidth]{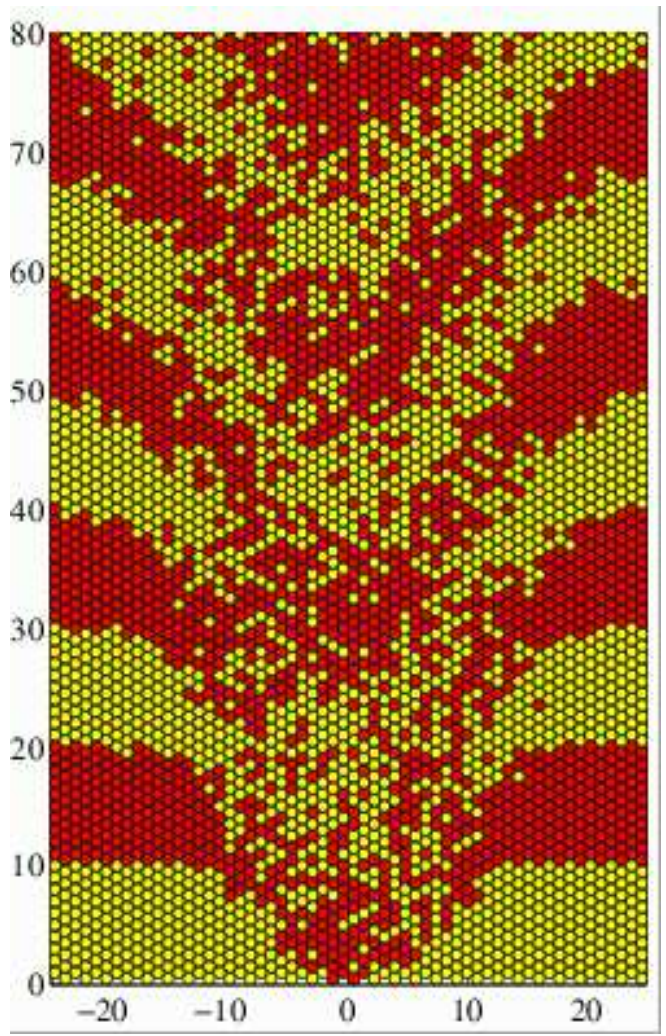} 
\caption{ \label{fig:void_layers} Simulation of the experiment in
Fig.~\protect\ref{fig:azadeh} using the Void Model on a triangular
lattice, which results in far too much mixing. (Courtesy of Chris
Rycroft.)   }
\end{center}
\end{figure}

To quantify the diffusion of particles, we consider geometrical
single-particle propagator, $\Pc_p(\xb|z,\xb_0,z_0)$, the conditional
probability density of finding a particle at horizontal position,
$\xb$, after it has fallen to a vertical position, $z$, from an
initial position, $(\xb_0,z_0)$.  For a steady mean concentration of
voids on a lattice, $\rho_v(\xb,z)$, it is straightforward to take the
continuum limit of the exact lattice dynamics (as will be described
elsewhere) to obtain a partial differential equation for
$\Pc_p(\xb|z,\xb_0,z_0)$ at length scales much larger than the
particle diameter:
\begin{equation}
- \frac{\partial \Pc_p}{\partial z} = b\, \del_\perp \cdot \left(
\del_\perp \Pc_p 
+ \Pc_p \, \del_\perp \log \rho_v \right).  \label{eq:Pcp}
\end{equation}
This equation may also be derived heuristically as follows. In a
uniform flow ($\rho_v =$ constant), a particle does a directed random
walk downward with precisely the same diffusion length as the voids
going upward. As in Eq.~(\ref{eq:rhov}) for the void concentration,
this yields the first two terms in Eq.~(\ref{eq:Pcp}), i.e. the
diffusion equation with ``$-z$'' acting like ``time''. In a nonuniform
flow, the third term represents a horizontal bias of the particle
random walk to climb the horizontal gradient of the void
concentration, which ensures that the propagator is eventually
localized on the source of voids (the silo orifice). The ``advection
velocity'', $b \del_\perp \log \rho_v$, is given by the ratio of (minus) the
horizontal void flux to vertical void flux, $b\del_\perp \rho_v / \rho_v$,
according to the right hand side of Eq.~(\ref{eq:voidflux}).  The task
of characterizing solutions to the new equation~(\ref{eq:Pcp}) is left
for future work.

\begin{figure*}
\begin{center}
\includegraphics[width=2.5in]{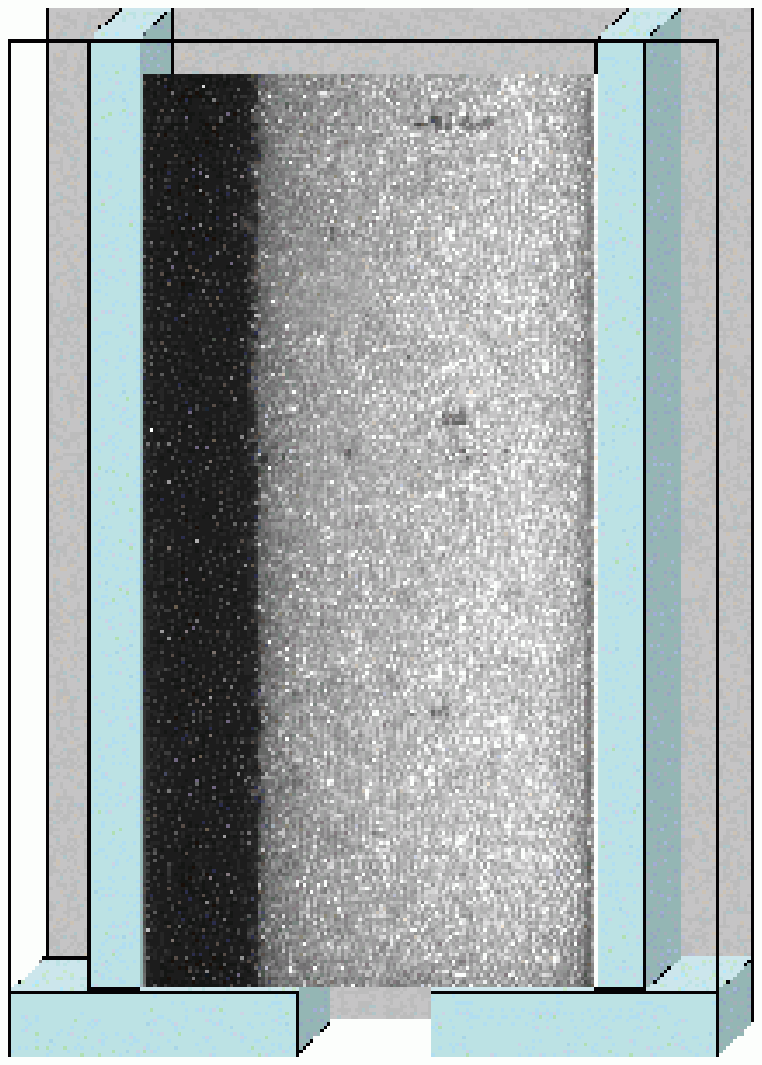} \nolinebreak
\includegraphics[width=1.8in]{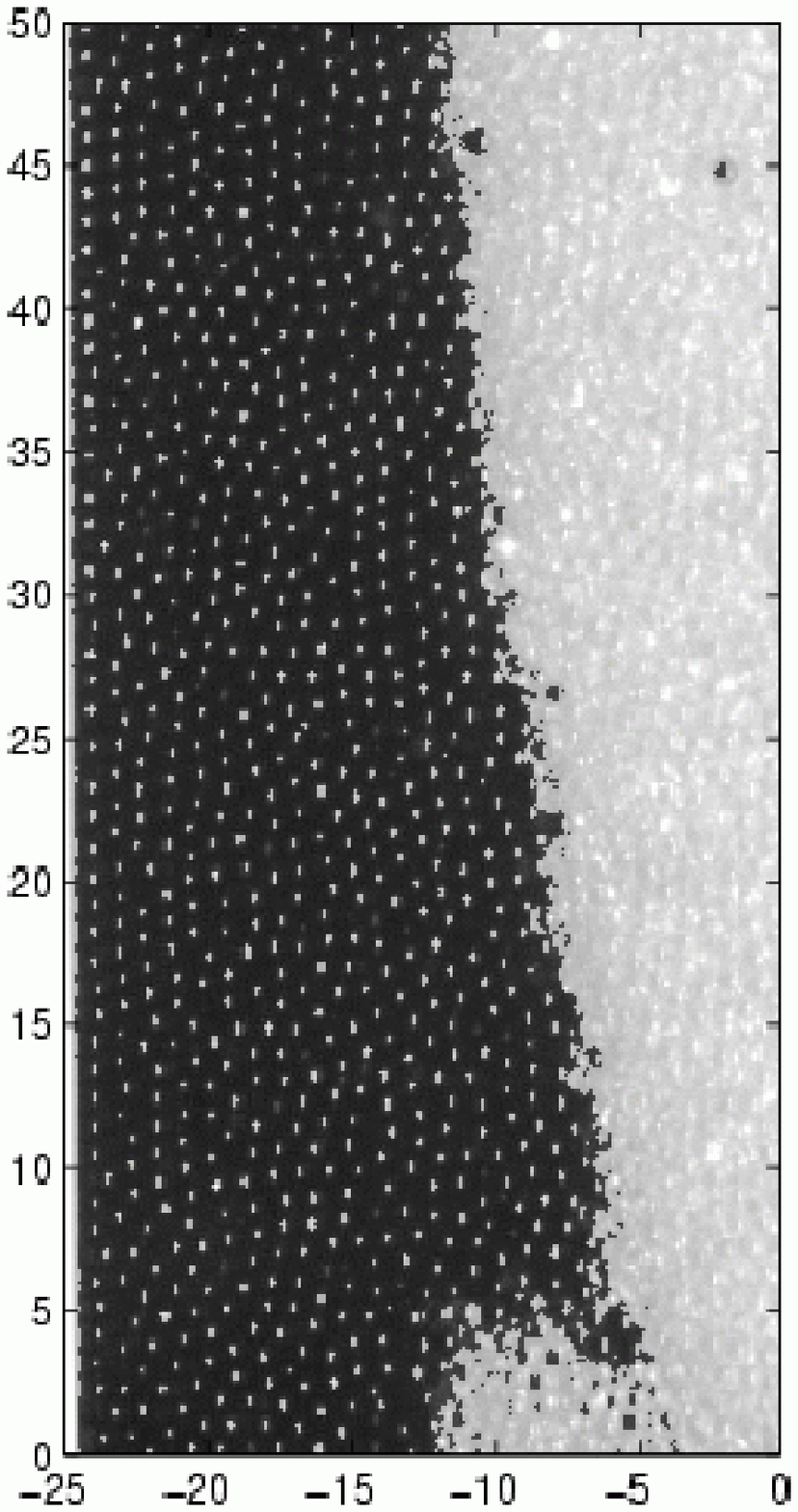}  \ \nolinebreak
\includegraphics[width=1.8in]{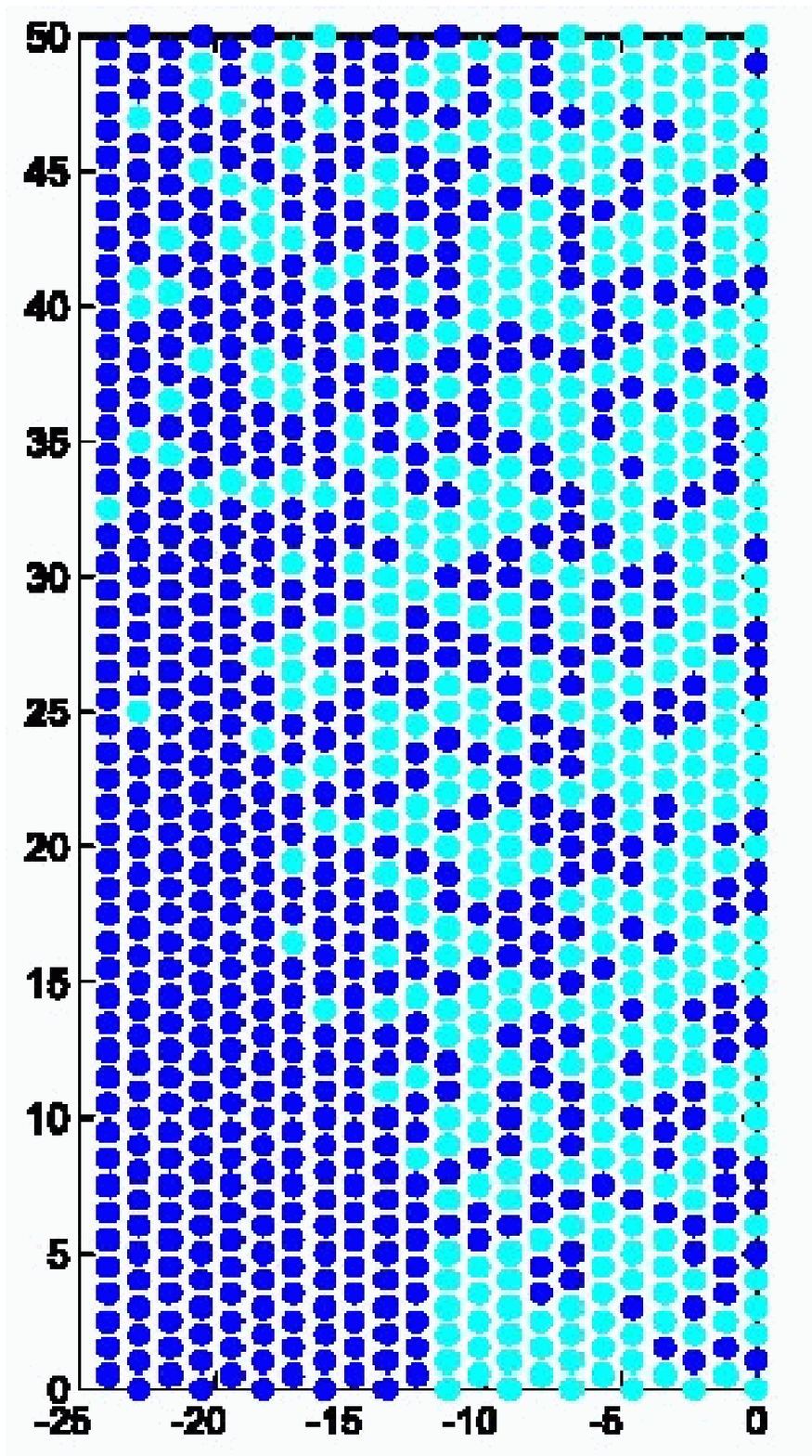}
\caption{ \label{fig:interface} Left: Experimental photograph of the
initial condition of a quasi-two-dimensional silo with a vertical
interface between black and white glass beads (by Jaehyuk
Choi~\protect\cite{choi04}). Center: The interface (in the left half
of the silo) at a later time in the experiment. Right: A simulation of
the same situation using the Void Model on a square lattice (by
Gu\'aqueta~\protect\cite{camilo03}). Length is measured in particle
diameters ($d=3$mm). }
\end{center}
\end{figure*}

\subsection{ The Paradox of Granular Diffusion }

Here, it suffices to note that the particle dynamics in the Void Model
is firmly contradicted by experiments. In a uniform flow,
Equation~(\ref{eq:Pcp}) predicts that particles follow statistically
identical geometrical trajectories while falling as voids do while
rising, with the same diffusion length, $b$. Figure~\ref{fig:azadeh}c
provides an elegant visual demonstration that this prediction is
incorrect.  The Void Model would predict that a particle starting near
the top of the silo would diffuse horizontally within an inverted
parabolic region of the same curvature as the parabolic flow region
(set by the diffusion of voids). Therefore, a collection of particles
starting near the top of the silo inter-diffuses and mixes across most
of the flow region before falling to the middle of the silo, as shown
in Fig.~\ref{fig:void_layers}. Instead, experimental particles mix
much less while draining, even in the non-uniform flow region, as the
initially sharp interfaces between layers of different colors remain
relatively intact during drainage.

The failure of the Void Model has a clear microscopic
origin. Particle-tracking experiments on silo drainage (motivated by
the theory below) have recently shown that the "cage" of neighbors of
a particle is preserved over surprisingly long distances, on the order
of hundreds of particle diameters, even comparable to the system
size~\cite{choi04}. This experimental result violates the central
hypothesis of the Void Model, illustrated in Fig.~\ref{fig:void}, in
any of its forms. On a lattice~\cite{hong91}, when a particle moves by
interchanging with a diffusing void, it loses roughly half of its
neighbors. As this process is repeated, the particle is likely to lose
all of its original neighbors after falling by only a few times its
own diameter. This quickly results in a rather unphysical degree of
mixing, as shown in Figs.~\ref{fig:void_layers} and ~\ref{fig:interface}.

This same conclusion also applies to the early engineering theories,
which are more vague about the microscopic
dynamics~\cite{nedderman}. Litwiniszyn~\cite{lit58,lit63a,lit63b}
postulated that particles jump downward from one fixed cage to
another, but this implies that a particle completely breaks free of
its neighbors with each random displacement. Similarly,
Mullins~\cite{mullins72,mullins74,mullins79} postulated that
individual particles move by exchanges with particle-sized voids,
independently from the random motions of neighboring particles, and
this again implies a cage-breaking length comparable to the particle
diameter.  In reality, particles diffuse in a cooperative fashion,
somehow managing to preserve their cages of first neighbors over
rather large distances.

How is it possible that the Void Model is able to describe the
macroscopic flow while so poorly describing the microscopic dynamics?
This illustrates the danger of judging a statistical hypothesis based
on only mean quantities, as some authors have argued that the success
of the Kinematic Model lends support to the Void Model. The ``paradox'' of
dense granular flow is that the diffusivity of particles is several
orders of magnitude smaller than the diffusivity of free volume.

\subsection{ More Clues from Experiments }

A closer look at experimental images suggests the key to resolving
this paradox.  As shown in fig.~\ref{fig:interface}(a), an interface
between differently colored (but otherwise identical) particles
deforms considerably in a non-uniform, dense granular flow, and yet
individual particles rarely (if ever) penetrate from one region into
the other. This is in stark contrast to diffusion in normal liquids
and gases, where the independent random walks of tiny molecules cause
an initially sharp interface to become smoothly blurred over time,
even in the absence of a flow. Instead, we see that the granular
interface in the center panel of Fig.~\ref{fig:interface} develops
some waviness at the scale of several particle diameters. This
suggests that particles somehow diffuse {\it cooperatively} in
cage-like blocks, mostly staying together with their nearest
neighbors.

This intuitive notion is consistent with other experimental
indications of an important length scale of several particle
diameters, which is missing in the Void Model. It is well known that a
silo will not begin to drain until the orifice is at least several
particles wide, and mechanical blocking (due to arching) can occur at
somewhat larger orifice widths~\cite{nedderman}. The empirical
Beverloo correlation~\cite{beverloo61} implies that the extrapolated
outlet diameter where the flow-rate vanishes is roughly 1.5 particle
diameters, which has been explained in terms of a controversial empty
annulus near the edge of the orifice~\cite{brown}. For our purposes,
the salient point is that a granular material does not drain one
particle at a time, so it is impossible to inject individual voids at
the orifice. Since packing constraints must be enhanced in the bulk
compared to the orifice, it seems highly improbable that voids could
form and propagate in the interior of the silo.

On the other hand, if particles tend to preserve their cages while
diffusing in the bulk, then they should only pass through the orifice
in correlated groups. Particle-tracking measurements of silo drainage
show some signs flow intermittency until the opening is at least six
particles wide~\cite{choi04}, consistent with the point where the
Beverloo correlation becomes acceptable~\cite{nedderman}. This again
suggests that extended blocks of particles must be allowed leave the
silo at the same time. Thinking in terms of the Void Model, it could
also mean that some extended entity causing motion is released into
the bulk, once the orifice reaches a critical width.

This idea may seem strange, but it has a chance of being correct.  The
Void Model does reproduce one crucial feature of particle dynamics in
slow, dense drainage experiments~\cite{choi04}: All fluctuations are
independent of the flow rate, aside from a simple rescaling of time.
One implication is that the P\'eclet number,
\begin{equation}
\Pe = \frac{U d}{D}, \label{eq:Pe}
\end{equation}
is independent of the flow rate, or equivalently the diffusivity, $D$,
is proportional to the local mean velocity, $U$.
This strongly suggests that {\it diffusion and advection are caused by
the same physical mechanism}, as in the Void Model. The common
mechanism, however, cannot be the propagation of a void or a
cage-breaking event, which implies $\Pe \approx 1$. Instead, it seems
to involve the cooperative motion of neighboring particles, which
somehow results in much less diffusion, $\Pe \gg 1$.

\section{ The Spot Hypothesis }
\label{sec:spot}

\subsection{ A Mechanism for Cooperative Diffusion }

In developing a microscopic theory, we are constrained by the fact
that the mean velocity is fairly well described by the Kinematic
Model, which seems to imply the diffusion of some entity causing
motion. Let us suppose that such an entity exists, but, since it
cannot be a void, we will call it a ``spot''. The kinematic parameter
is then approximately set by the spot diffusion length,
\begin{equation}
b = \frac{\var(\Delta\xb_s)}{2d_h\Delta z_s}  \label{eq:bspot}
\end{equation}
where $\Delta\xb_s$ is the random horizontal displacement of a spot as
it rises by $\Delta z_s$.

\begin{figure}[t]
\begin{center}
\includegraphics[width=2.7in]{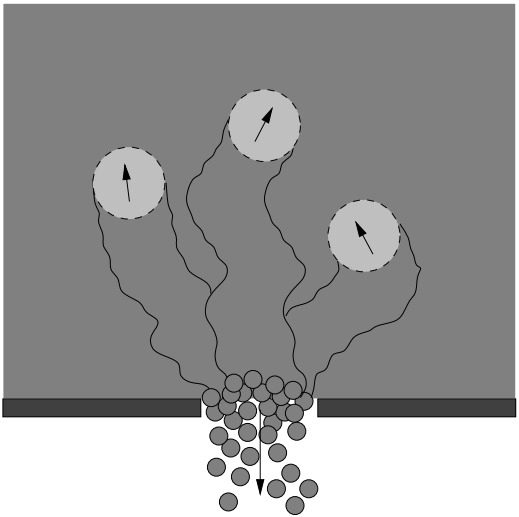}
\caption{ \label{fig:rising_spots} The Spot Hypothesis I: Localized
spots free interstitial volume, extending several particle diameters,
diffuse upward from the silo orifice as particles drain out, causing
the reverse reptation of thick chains of particles. }
\end{center}
\end{figure}

What exactly is a spot?  Since spots are injected at the silo orifice
as particles drain out, it is natural to assume that each carries a
certain amount of free volume. Rather than being concentrated in a
void, however, this volume should correspond to {\it a slight excess
of interstitial space spread across an extended region}, typically
larger than a particle, as shown in Fig.~\ref{fig:rising_spots}. We
expect the size of a spot to be at least three particle diameters
since it should induce the cooperative motion of a particle and its
nearest neighbors.

\begin{figure}[t]
\begin{center}
\includegraphics[width=2.7in]{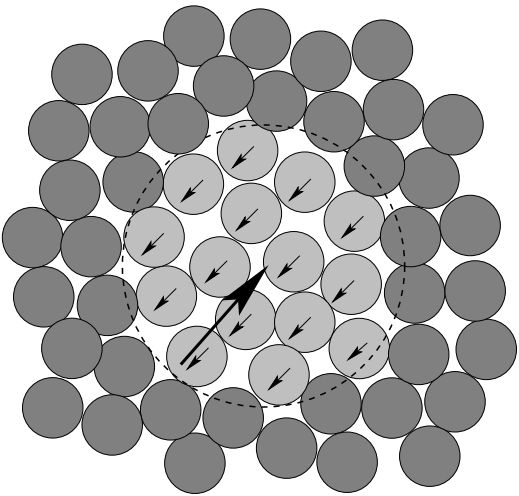}
\caption{ \label{fig:spot} The Spot Hypothesis II: Each spot
displacement causes a block of neighboring particles to make a smaller
displacement in the opposite direction, so as to roughly conserve the
local volume.  }
\end{center}
\end{figure}

The simplest dynamical hypothesis, illustrated in Fig.~\ref{fig:spot},
is that {\it a spot causes all affected particles to move as a block}
with the same small displacement in the opposite direction. As the
spot follows its random trajectory upward, a thick chain of particles
cooperatively reptates downward by a very small distance, much less
than the particle diameter.  The subsequent passage of other spots
causes each particle to randomly reptate in many different thick
chains, as shown in Fig. ~\ref{fig:rising_spots}.

Under this dynamics, the packing of particles is largely preserved
because a pair of neighboring particles is usually affected by the
same spot. Occasionally the pair finds itself near the edge of a spot,
which causes a tiny relative displacement. The accumulated effect of
such events gradually causes the particles to separate as they
drain. The cage breaking length, however, is much larger than the
particle size because relative displacements of neighbors are both
small and rare. In fact, the cage-breaking length in the model
(calculated below) can be comparable to the system size, as in
experiments and simulations of dense drainage.

This microscopic picture is fundamentally different from previous
theories, which assume that particles undergo independent random
walks. According to Kinetic Theory, a particle moves ballistically
between instantaneous, randomizing collisions, and in the Void Model it jumps from
cage to cage. In both cases, the particle loses a most of its
neighbors in a single displacement, and thorough mixing occurs at the
scale of several particle diameters.

\subsection{  Correlations Reduce Diffusion }

A back-of-the-envelope calculation suffices to show that the spot 
hypothesis resolves the paradox of granular diffusion.  Suppose that a spot
carries a total free volume, $V_s$, and causes equal displacements,
$(\Delta \xb_p,\Delta z_p)$, among $N_p$ particles of volume,
$V_p$. The particle displacement can be related to the spot
displacement, $(\Delta \xb_s,\Delta z_s)$, by an approximate
expression of total volume conservation,
\begin{equation}
N_s \, V_p \, (\Delta \xb_p,\Delta z_p) = - V_s\, (\Delta \xb_s,
\Delta z_p) , \label{eq:spotsimple}
\end{equation}
which ignores boundary effects at the edge of the spot. From this
relation, we can easily compute the particle diffusion length,
\begin{equation}
b_p = \frac{\var(\Delta\xb_p)}{2d_h|\Delta z_p|} =
 \frac{w^2 \var(\Delta\xb_s)}{2d_h w \Delta z_s} = w\, b  \label{eq:bpsimple}
\end{equation}
which is smaller than the spot (or volume) diffusion length by a
factor,
\begin{equation}
w  = \frac{V_s}{N_p V_p} ,  \label{eq:wsimple}
\end{equation}
equal to the ratio of the spot's free volume, $V_s$, to the total
affected particle volume, $N_p V_p$. The Void Model corresponds to the
unphysical limit where these volumes are both equal to a single
particle volume ($N_p=1, V_s=V_p, w=1$), but generally we expect spots
to affect multiple particles and carry relatively little excess volume
($N_p \gg 1, V_s < 1, w \ll 1$). The nontrivial implication of
cooperative motion is then that particles diffuse much more slowly
than volume, $b_p \ll b$, which resolves the paradox
described above. Below we will show that the resolution is not just
qualitative, but quantitative.

\subsection{ General Formulation of the Spot Model }

The key mathematical concept in the Spot Model is that of a diffusing
``region of influence'', which causes correlated displacements among
all particles within range.  There are many possible microscopic
postulates for this local cooperative motion, but the strong tendency
to preserve nearly jammed packings suggests it should be mainly a
block translation as described above, plus perhaps a random block
rotation, mostly involving nearest neighbors. For such rigid-body
motions, however, the shear at the edge of a spot may be somewhat
excessive.

More generally, we may allow the influence of a spot to be governed by
a smooth function, $w(\rb_p,\rb_s)$, which decays quickly with distance
from the spot center, $|\rb_p-\rb_s|$.  For example, in the case of
purely translational motion in one direction (opposing a 
spot displacement), the random 
displacement, $\Delta\Rb_p^{(i)}$, of the $i$th particle centered at
$\rb_p^{(i)}$, due to the random displacement, $\Delta \Rb_s^{(j)}$, of $j$th
spot cented at $\rb_s^{(j)}$ would be given by 
\begin{equation}
\Delta \Rb_p^{(i)} = - w(\rb_p^{(i)},\rb_s^{(j)}+\Delta\Rb_s^{(j)}) \,
\Delta\Rb_s^{(j)}  \label{eq:discrete}
\end{equation}
We might expect a spot to roughly maintain its ``shape'' while moving
through a system of particles, due to the local statistical regularity
of random packings, so its influence function, $w$, should depend
mainly upon the separation vector, $\rb_p^{(i)} - \rb_s^{(j)}$. This
may change in regions of highly non-uniform flow, but in
any case the spot influence should typically decay at separations,
$|\rb_p^{(i)}-\rb_s^{(j)}|$, larger than a few particle diameters.
Below we demonstrate that the mathematical model of
Eq.~(\ref{eq:discrete}) is a natural starting point for discrete
simulations, as well as analysis in the continuum limit. 

There are many possible extensions, which should be considered in
future work. For example, the internal particle motion induced by a
spot could contain a random component, or some other more complicated
cooperative motion driven by inter-particle forces (see
below). Different spots could also interact with each other by
changing their drift velocity, diffusion length, size and/or free
volume, which effectively induces cooperative particle motion at large
scales. Weak repulsion and attraction might both occur, since the
volume fraction tends to remain in a certain narrow range in a region
of flow. Even if spots remain unchanged during motion, there could be
statistical distributions of spot free volume, size, and shape, which
might vary due to compaction and dilation processes driven by random
spot annihilation, creation, and recombination. Such extensions may be
necessary for a complete theory of dense granular flow, but in the
next section we show that the basic model already compares well with
silo-drainage experiments.

\section{ Experimental Evidence  }
\label{sec:expt}

In this section, we consider the simplest version of the Spot Model,
in which each spot carries the same free volume, $V_s$, and is
spherical with a uniform influence, $w$, out to a distance $R$. We
compare this model to experimental data, primarily from
Ref.~\cite{choi04}, including some new results by the same authors.
The experiments provide compelling tests of the theoretical
predictions, because the latter were made well in advance~\cite{spot01}.

\subsection{ Free Volume and the P\'eclet Number }

The Spot Model makes a quantitative connection between dilation,
diffusion, and advection in a dense granular flow. As such, it
predicts the P\'eclet number (\ref{eq:Pe}) in terms of typical local
volume fluctuations.  The P\'eclet number is independent of the flow
rate, but, unlike with the Void Model, the predicted magnitude is
consistent with experimental data.

We start with the general observation that the particle volume
fraction, $\phi \approx 0.6$, varies by at most a few percent in the region of
dense flow, away from the orifice. This is certainly a reasonable
inference from drainage experiments, which maintain a nearly uniform
density to the naked eye~\cite{choi04}. Detailed observations of
particles near a glass silo wall confirm that the local area fraction
varies by one to three percent at different flow
rates~\cite{choi_note}. Likewise, in granular dynamics simulations,
the bulk volume fraction in thin horizontal cross sections varies in
the range, $\phi\approx 0.565-0.605$, near the orifice, while staying
within one or two percent of 0.60 in the region of steady dense flow
far above the orifice~\cite{landry_note}.

These values are also consistent with studies of random packings of
hard spheres in zero gravity. Rigidity percolation and dilatency
(expansion under shear stress) set in at $\phi \approx 0.55$ for
``random loose packing''~\cite{onoda90}, which provides a rough lower
bound on the volume fraction in a slow, dense flow. A rough upper
bound is set by jammed random packing~\cite{torquato}, e.g. in the
``maximally random jammed state'' ( $\phi \approx 0.64$
~\cite{torquato00} or $\phi\approx 0.63$~\cite{kansal02}) or the
zero-temperature, zero-stress ``jamming point'' ($\phi \approx 0.63$
~\cite{ohern02,ohern03}). Higher volume fractions also possible at the
expense of some long-range crystalline order (e.g. as demonstrated by
experiments with horizontal shaking~\cite{pouliquen97}), up to the
rigorous upper bound of $\phi = \pi/\sqrt{18} \approx 0.74$ for the
FCC lattice~\cite{hales04}, but such configurations are unlikely to
permit quasi-steady flow. 

If we attribute the maximum local reduction in volume fraction,
$\Delta \phi/\phi \approx 0.01$, to the presence of $N_s$ overlapping
spots,
\begin{equation}
\Delta\phi = \phi - \frac{\phi}{1 + \phi w N_s} \approx \phi^2 w\, N_s,
\end{equation}
then we have 
\begin{equation}
w N_s \approx \frac{\Delta\phi }{\phi^2 } \approx
\frac{0.01}{0.6} \approx 0.02
\end{equation}
In that case, the P\'eclet number 
for horizontal diffusion would be 
\begin{equation}
\Pe_x = \frac{ (|\Delta z_p|/\Delta
t) d}{ \var(\Delta\xb_p)/2 \Delta t}
 = \frac{d}{b_p} = \frac{d}{w b} \approx 40\, N_s   \label{eq:Pe_est}
\end{equation}
for $b=1.3 d$. Since $N_s \geq 1$ in this calculation, we predict that
a particle falls by on the order of 100 diameters before diffusing
horizontally by one diameter.

This simple estimate is consistent with the experimental value,
$\Pe_x = 321$, for particles near a smooth wall in slow drainage from
a quasi-two dimensional silo~\cite{choi04}.  It also suggests that
spots occur at fairly high density, with as many as $N_s = 320/40 = 8$
overlapping in a position of high dilatency.  In a position of low
dilatency, where $\Delta \phi/\phi \approx 0.001$, it implies that
spots typically do not overlap. Such a limited number of spot overlaps
seems like a reasonable definition of a ``dense" flow, where the simple
picture in Eq.~(\ref{fig:spot}) could apply. 

(Note that larger P\'eclet numbers in the range $\Pe \approx
1500-3000$ have been reported for faster flows in long, narrow silos
with shear-inducing rough walls~\cite{natarajan95}. In this case, the
simple relation in Eq.~(\ref{eq:Pe_est}) would imply a much higher
spot density, where positions of 100 overlapping spots could be
found. At such high spot densities (low volume fraction), it seems the
model would no longer apply due to more independent particle
displacements.  Indeed, since $\Pe$ depends weakly on the flow rate,
those experiments are not in the regime of slow drainage considered
here. The data is also inconsistent with kinetic theories of
dilute granular flow~\cite{natarajan95}, so it seems to correspond to
an intermediate regime of moderately dense flow.)

\subsection{ Spatial Velocity Correlations }

The quantitative resolution of the granular-diffusion ``paradox''
provides some support for the Spot Model does not directly validate
its microscopic hypothesis. How can we directly confirm or reject the
existence of spots?  The calculation above suggests that it would be
impossible to observe the propagation of a single spot, but only large
numbers of spots, which expectation is consistent with x-ray
diffraction experiments showing fairly smooth density patterns in
draining sand~\cite{behringer89}.

Rather than trying to observe a single spot, therefore, it makes more
sense to seek statistical evidence of the passage of many spots.  A
direct signature of the cooperative motion in Fig.~\ref{fig:spot} is
found in the spatial correlation function of velocity fluctuations
(relative to a steady mean flow).  Two particles are likely to
fluctuate in the same direction when they are separated by less than a
spot diameter because most spots engulf them at the same time. On the
other hand, more distant particles are always affected by different
spots, which implies independent fluctuations.

The spatial velocity correlation function, $C(r)$, for two particles
separated by $r$ is easily calculated for uniform, spherical spots of
radius, $R$. The two instantaneous particle displacements are either
identical (perfectly correlated), if they are caused by the same spot,
or independent.  Therefore, the correlation function, $C(r)$, is
simply the scaled intersection volume, $\alpha(r;R)$, of two spheres
of radius, $R$, separated by a distance, $r$:
\begin{equation}
\alpha(r;R) = 
1 - \frac{3}{4}\frac{r}{R} + \frac{1}{16}\left(\frac{r}{R}\right)^2 
.
\label{eq:alpha}
\end{equation}
This result appears in a recent study of random point
distributions~\cite{torquato03}, where $\alpha(r;R)$ is plotted in
Fig. 9 (and earlier in Ref.~\cite{torquato85}). The key point is that
$C(r) = \alpha(r;R)$ decays to half its maximum value at a separation
slightly smaller than spot radius, $R$.  Similar curves result
from other quickly decaying spot influence functions, such as
exponentials or Gaussians (see below).

\begin{figure}
\begin{center}
\includegraphics[width=\linewidth]{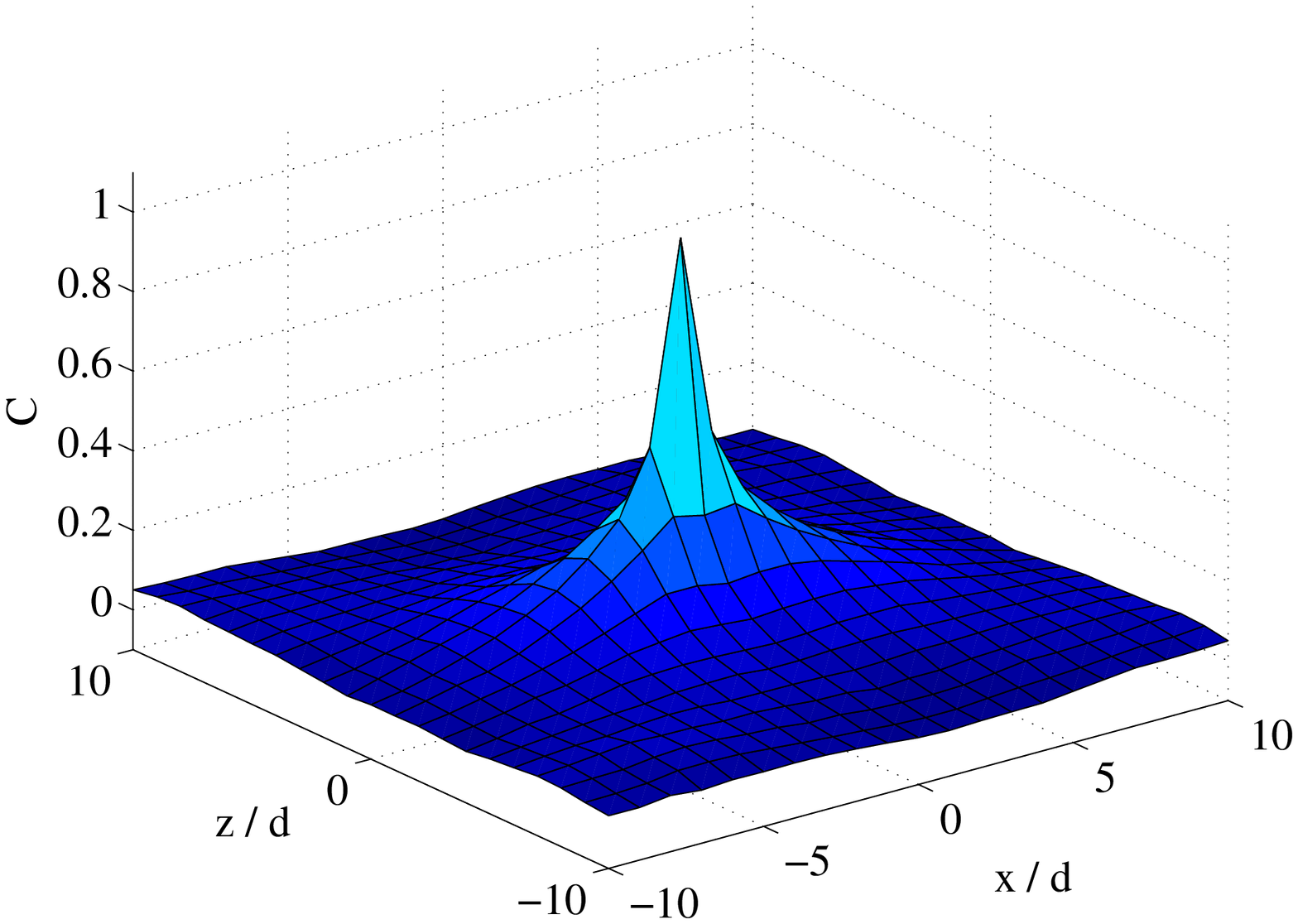}\\
\includegraphics[width=\linewidth]{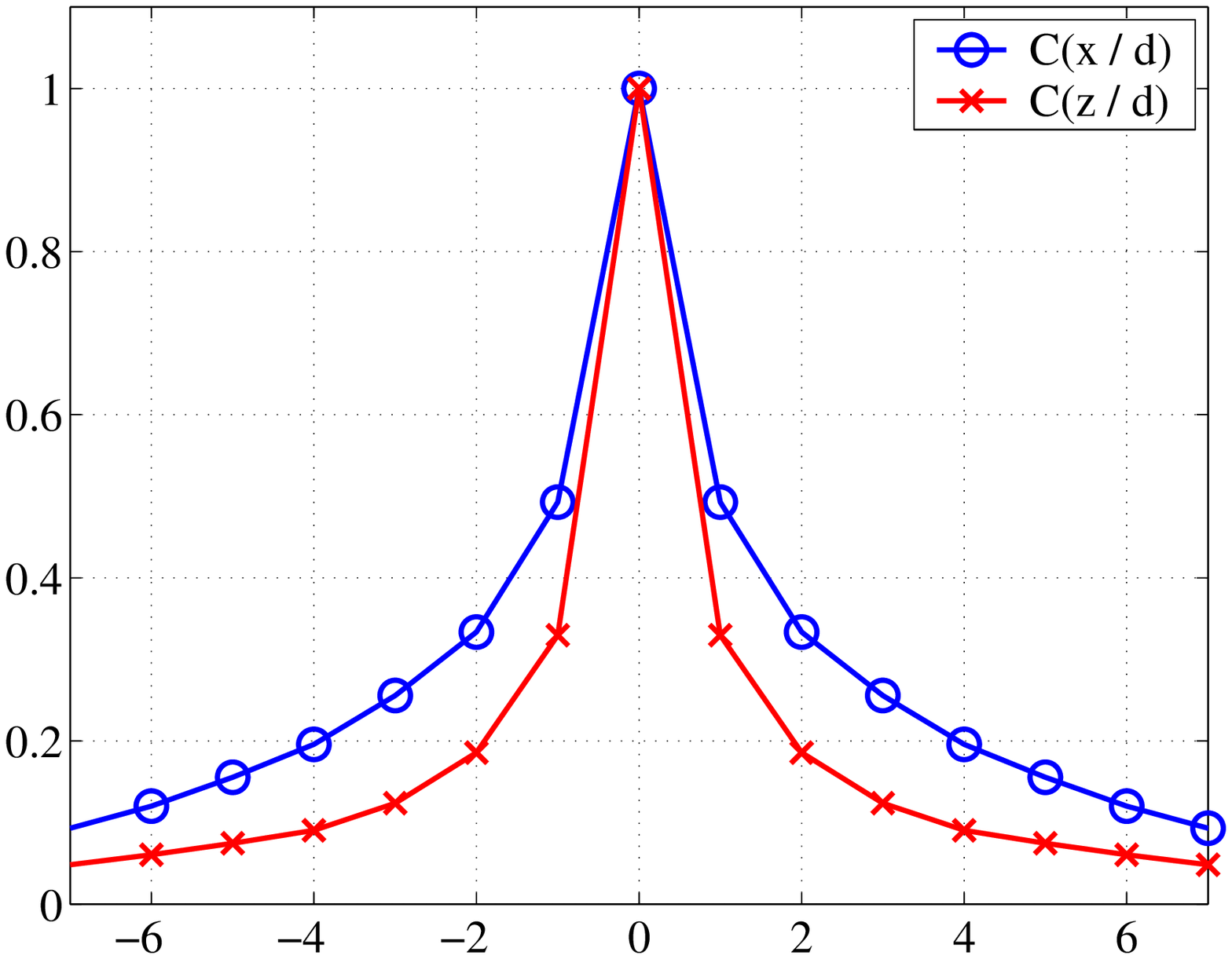}
\caption{ \label{fig:Corr} Direct experimental evidence for spots:
Spatial velocity correlations from the drainage experiments of
Ref.~\cite{choi04}.  Surface plot
(top) and coordinate slices (bottom) of the correlation coefficient
for velocity fluctuations (about the local mean flow) at horizontal
separation, $x$, and vertical separation, $z$, in units of particle
diameter, $d$.  (Courtesy of Jaehyuk Choi.)}
\end{center}
\end{figure}

Motivated by this prediction, the experiments in Ref.~\cite{choi04}
were performed to look for spatial velocity correlations in a real
granular flow.  Hundreds of glass beads ($d=3$mm) near the wall of a
quasi-two-dimensional draining silo were simultaneously tracked with
$1$ms resolution in time and $d/100$ in space in the region of
non-uniform flow near the orifice. The data clearly reveals spatial
velocity correlations, shown in Fig.~\ref{fig:Corr}, at the expected
length scale of several particle diameters.  It is interesting that
real spots are not quite spherically symmetric. The velocity
correlations extend roughly twice as far in the horizontal direction
as in the vertical direction, so these spots are more elliptical in
shape.  The data also implies a smooth decay of the spot influence
function, $w(\rb_p,\rb_s)$, roughly an exponential decay. In contrast,
for a uniform spot influence with a cutoff radius, $R$, the
correlation function, $C(r) = \alpha(r,R)$, would decay roughly
linearly to zero at $r=2R$.

Leaving these interesting details for future analysis, we proceed with
our assumption of uniform spot influence, as a good first
approximation. As expected for cooperative motion largely preserving
particle cages, the spot diameter is roughly five particle diameters,
which corresponds to tens of affected particles per spot (say $N_p
\approx 20$). Using Eq.~(\ref{eq:Pe_est}) to infer spot influence,
\begin{equation} 
w = \frac{d}{\Pe_x b} =
\frac{1}{(321)(1.3)} = 0.0024 , \label{eq:wexpt} 
\end{equation} 
the
simple relation, Eq.~(\ref{eq:wsimple}), then implies,
\begin{equation} 
\frac{V_s}{V_p} = N_p w \approx 0.05 
\end{equation}
In other words, {\it the interstitial free volume carried by a spot,
although spread across tens of particles, is only a few percent of one
particle volume}. Such small, coherent volume fluctuations are
completely different from those postulated by the Void Model
($V_s=V_p$, $R=R_p$).

\subsection{ Cage Breaking }

Further support for the spot hypothesis comes from experimental data
on cage breaking~\cite{choi04}, which also provides the most direct
contradiction of the Void Model. Visual observation of particles in
draining silos reveals that particle cages tend to persist for large
distances, comparable to the system size.  It is not unusual for some
first neighbors to remain unchanged from the middle of the silo all
the way to the orifice. A detailed analysis of the uniform-flow region
far from the orifice yields a cage-breaking length, $z_c \approx 200
d$, extrapolated from small falling distances, $\Delta z\leq 5 d$. As
such, the precise value, $200 \,d$, should not be taken too seriously,
but it is clear that cage breaking occurs very slowly in regions of
low shear, at scales comparable to the system size. In contrast, the
Void Model inevitably predicts cage-breaking at the scale of a single
grain, $z_c\approx d$.

Rather than analyzing many-body cage dynamics in the Spot Model, it is
simpler to consider two tracer particles, which are initially first
neighbors, in a nearly uniform flow. Without correlations, the 
horizontal separation, $X$, would grow with distance dropped, $z$,
like $X \approx d + \sqrt{2 b_2 z}$, where the diffusion
length for relative motion is twice that of single particles, $b_2
= 2 b_p$, due to the additivity of variance. With correlations,
the relative diffusion length,
\begin{equation}
b_2(X) = 2\, b_p \left[ 1- C(X) \right] ,
\end{equation}
is reduced by a factor, $1 - C(X) \approx d/2R$ (for $X\approx d$).
Supposing that cage breaking occurs at $X \approx 2 d$, we predict a
cage-breaking length, $z_c = d^2/2b_2$, of roughly,
\begin{equation}
\frac{z_c}{d} \approx \frac{R}{2 b_p} = \frac{\Pe_x R}{2\, d } \approx
\Pe_x
\end{equation}
where we take the approximation, $R\approx 2d$, from the data in
Fig.~\ref{fig:Corr}.  The relation, $z_c \approx \Pe_x d \approx 300
d$, is quite consistent with the experimental data.

\subsection{ Vertical Diffusion }

We have postulated that spots diffuse upward with a horizontal
diffusion length equal to the kinetic parameter, $b_s^\perp = b$,
because the mean velocity roughly follows the profile of the Kinematic
Model, as shown in Fig.~\ref{fig:azadeh}. By looking at fluctuations,
we can also infer the vertical diffusion length of spots, which would
be difficult to measure directly in experiments. (Note that voids
exhibit zero vertical diffusion, unless one introduces a random
waiting time between steps.)

Suppose that the vertical component of the spot displacement, $\Delta
z_s$, is a positive random variable with a finite
mean,  $\langle \Delta z_s \rangle$.  We can then define the vertical
diffusion length of spot,
\begin{equation}
b_s^\| = \frac{ \var(\Delta z_s) }{2 \, d_h \langle \Delta z_s \rangle},
\end{equation}
by analogy with Eq.~(\ref{eq:bspot}) for the horizontal diffusion
length (where $\Delta z_s$ should be replaced by $\langle \Delta z_s
\rangle$).  In a similar way, we define the vertical diffusion
length of particles, $b_p^\|$ (and $b_p^\perp = b_p$).

Assuming that the spot influence  does not distinguish between different 
velocity components, as in Eq.~(\ref{eq:spotsimple}), we have the
identity,
\begin{equation}
w = \frac{b_p^\|}{b_s^\|} = \frac{b_p^\perp}{b_s^\perp} . \label{eq:bw}
\end{equation}
This allows us to infer the vertical diffusion length of spots from
that of particles,
\begin{equation}
b_s^\| = \frac{b_p^\|}{w} = \frac{d}{\Pe_z w} =
\frac{d}{(150)(0.0024)} = 2.8\, d  \label{eq:bs}
\end{equation}
where we take $\Pe_z=150$ from experiments~\cite{choi04}. This
calculation assumes that the dominant contribution to the particle
vertical diffusivity is the influence of the spot vertical
diffusivity, and not the random arrival of different spots (consistent
with the mean-field analysis of section~\ref{sec:math}). Since
$b_s^\|/b_s^\perp = 2.8/1.3 \approx 2$, the spot displacements in these
experiments fluctuate twice as much in the vertical direction as in
the horizontal direction.

\subsection{ Spot Interactions and Density Waves }

Until now, we have assumed that spots undergo independent random
walks. This suffices to describe many aspects of silo drainage, so we
conclude that spot interactions are relatively weak in such
flows. Nevertheless, there must be short-range interactions between
spots to keep the local volume fraction in the typical range for
dense, gravity-driven flows ($0.56 \leq \phi \leq 0.61$).  Physically,
spot repulsion should derive from the tendency for particles to
quickly collapse from all directions toward a region of overly low
volume fraction ($\phi < 0.56$). Because each spot carries very little
volume, the repulsion should only become strong when many spots
overlap (e.g. $N_s > 10$). Similarly, spot attraction follows from the
tendency of free volume to avoid nearly jammed regions ($\phi >
0.61$) if it can be accommodated in an already flowing region. This
attraction should strengthen as spots become more dilute ($N_s \leq 1$),
e.g. sharpening the boundary between flowing and stagnant regions in
Fig.~\ref{fig:azadeh}.

\begin{figure}
\begin{center}
\includegraphics[width=2.8in]{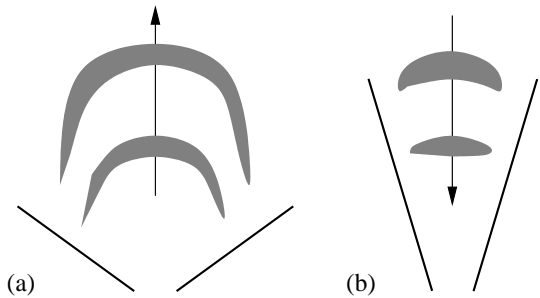}
\caption{ \label{fig:waves} Sketch of nonlinear waves of free volume
(gray regions) observed in funnels of draining
sand~\cite{behringer89}. For the same orifice width, the waves 
propagate upward for large opening angles (a) and sometimes downward
for small opening angles (b).  }
\end{center}
\end{figure}

Is there any evidence for spot interactions in experiments? An
indirect sign could be the density waves observed by Baxter {\it et
al.} in three-dimensional draining funnels of rough sand using x-ray
diffraction~\cite{behringer89}. The authors gave a heuristic
explanation of such waves in terms of the formation and collapse of
arches across the central region but did not attempt a quantitative
theory. As sketched in Fig.~\ref{fig:waves}, the observed density
patterns were mostly consistent with upward moving waves of reduced
density, but downward moving waves were also inferred in some narrow
funnels. Nonlinear wave equations for dilute flows~\cite{valance98}
have been derived from the kinetic theory of inelastic
gases~\cite{jenkins83}, but there does not appear to be a microscopic
theory for density waves in dense flows.

Subsequent authors~\cite{lee94a,lee94b,ristow94} have suggested an
analogy with nonlinear kinematic waves in traffic flow~\cite{whitham},
and some have even claimed that granular flow and traffic flow are in
the same ``universality class''~\cite{hayakawa98}. The classical,
continuum theory of traffic waves describes the effective short-range
repulsion between cars trying to avoid an accident by a flux-density
relation, $q(\rho)$, with a maximum at $\rho_c$.  The conservation law
for the car density,
\begin{equation}
\frac{\partial \rho}{\partial t} + \frac{\partial q}{\partial z} =
\frac{\partial \rho}{\partial t} + c(\rho)\frac{\partial\rho}{\partial
x}=0,
\end{equation}
then involves the characteristic velocity, $c(\rho) = q^\prime(\rho)$,
which changes sign at $\rho_c$, implying that density waves propagate
forward in light traffic ($\rho<\rho_c$) and backward in heavy traffic
($\rho>\rho_c$). Whenever density variations are generated (by noise,
boundary conditions, etc.), they combine into coherent structures
terminated on one side by a shock and on the other by an expansion
fan.

Granular shock waves in two-dimensional inclined funnels and pipes
have been studied extensively in
simulations~\cite{lee94a,lee94b,poschel94} and particle-tracking
experiments~\cite{horluck99,horluck01}. A detailed study of shock
statistics~\cite{horluck01} has revealed differences with simple
traffic-flow models, although somewhat similar average $q(\rho)$
curves have been measured. Granular dynamics simulations also show
that friction among the particles and walls initiates the fluctuations
leading to shock formation~\cite{ristow94}.

The Spot Model provides a simple microscopic explanation for the
experimental observations.  Given our expectation of weak spot
repulsion, it is natural to think of spots as ``cars'' moving upward
in a dense flow.  For a given orifice width, a funnel with a large
opening angle, as in Fig.~\ref{fig:waves}(a), should have a fairly low
spot density as a result of lateral diffusion. In that case ($\rho_s <
\rho_c$), nonlinear spot waves initiated by fluctuations at the
orifice propagate upward at nearly the spot drift velocity, slowly
organizing into coherent structures of reduced density. The backside
of the wave forms shock, below which the particle density suddenly
increases. Particles gradually accelerate into the looser spot-rich
region and suddenly slow down upon entering the more compact region
below.

As the funnel angle is reduced (for the same orifice width), spots are
increasingly reflected by the boundaries into the central region,
where their density goes up. The crowding of spots could eventually
cause a reversal of density waves, analogous to the familiar
backward-moving waves of heavy traffic. In that case ($\rho_s >
\rho_c$), a region of reduced spot density would form a shock on its
back (upper) side, which the particles would see as downward moving
compaction wave. On the other hand, a region of increased spot density
would form a shock on its front (lower) side, corresponding to a
downward moving rupture zone.

It would be interesting to revisit experimental density patterns in
light of the Spot Model. Density waves in wide funnels ($\rho_s <
\rho_c$) should move as weakly interacting ``tracer spots'', thus
revealing the spatial profile of spot drift and diffusivity. For
example, the wave patterns sketched in Fig.~\ref{fig:waves}(a) suggest
that the spot drift velocity and diffusion length are smaller near the
stagnant region than in the faster-flowing central region.

\subsection{ Crossover from Super to Normal Diffusion }

We have already mentioned the surprising result of particle tracking
in draining silos that all fluctuations depend on the distance
dropped, not time, over a wide range of flow rates~\cite{choi04}.  For
the mean-squared displacement relative to the mean flow, there is a
universal crossover from super-diffusion to normal diffusion after a
particle falls by roughly one diameter. The observed super-diffusive
regime, $0.005 d < |\Delta x| < 0.5 d$, is not ballistic ($|\Delta x|
\propto t \propto z$) but instead exhibits a non-trivial scaling
exponent, $\Delta x \propto t^{0.75} \propto z^{0.75}$.

A small-scale ballistic regime has also been inferred by
diffusing wave spectroscopy for glass beads ($d=95 \mu$m, $v=0.32$
cm/s) flowing down a vertical pipe toward a wire mesh~\cite{menon97}.
The data does not seem of indicative of thermal collisions, however,
since the ``collision distance'', $l_c = 28$ nm $\approx 10^{-5} d$,
is surely at the scale of surface roughness. Moreover, the ``collision
time'', $\tau_c = 9\,\mu$s $\approx 10^{-4} d/v$, implies a typical
velocity, $l_c/\tau_c = 0.31$ cm/s, equal to the mean flow speed,
which seems hard to attribute to independent thermal fluctuations. In
any case, no transition to diffusion is observed at this scale, so
these nano-collisions are not of the randomizing type postulated by
kinetic theory and have little bearing on our discussion of structural
rearrangements.

Returning to the particle-tracking data, we have seen that the Spot
Model describes the diffusive regime with quantitative accuracy, but
the super-diffusive regime is not so easily explained.  Spots are not
constrained to move on a lattice, so the size of the spot
displacements is arbitrary, as long as the horizontal spot diffusion
length equals the kinetic parameter, $b_s^\perp = b \approx d$. For
example, each spot could follow a persistent random walk in which
successive, small random displacements ($\Delta z_s \ll d$) have a
correlation coefficient, $\gamma$. Following
Taylor~\cite{taylor1921,weiss02}, the kinetic parameter would then be
\begin{equation} 
\frac{b}{b_0} = \frac{1+\gamma}{1-\gamma}  \label{eq:bper}
\end{equation} 
where $b_0$ is the spot diffusion length in the absence of
auto-correlations, given by Eq.~(\ref{eq:bspot}).  For $0<\gamma<1$,
the persistent random walk of the spot has ballistic scaling up to a
crossover distance, $-\Delta z_s/\log \gamma$, which should be of
order $d$, since $b \approx d$.  Since the spot diameter is larger
than $d$, a single particle is affected many times by the same spot,
so the particle would inherit a persistent random walk with ballistic
motion up to a crossover distance of roughly $w\, d \ll d$.

This scaling and crossover distance do not agree with experiment, but
various modifications could perhaps resolve the discrepancy. It is
straightforward to achieve sub-ballistic scaling with longer-range,
power-law auto-correlations between spot steps, although some cutoff
would be needed to preserve the normal diffusive regime.  More likely,
however, the super-diffusive regime is associated with more complicated
particle motion caused by spots. The simple picture in
Fig.~\ref{fig:spot} ignores packing constraints and frictional
contacts, so we should expect additional microscopic rearrangements to
occur, superimposed on the mean spot-induced motion. This fundamental
issue becomes more clear in simulations of the model.

\section{ Spot-Based Simulations }
\label{sec:sim}

\subsection{ Statistical Dynamics of Dense Flow }

The Spot Model provides a very simple and efficient Monte Carlo
algorithm for dense granular drainage. The simulation begins with a
given distribution of tracer particles, either on a lattice or in a
random packing. The particles then move passively in response to spots
undergoing directed random walks upward. Spots are injected randomly
in time and space along the orifice to artificially set the flow
rate. (Our theory makes no prediction about how spots are created,
which is related to the poorly understood connection between flow rate
and orifice width.)

\begin{figure}
\begin{center}
\includegraphics[width=3in]{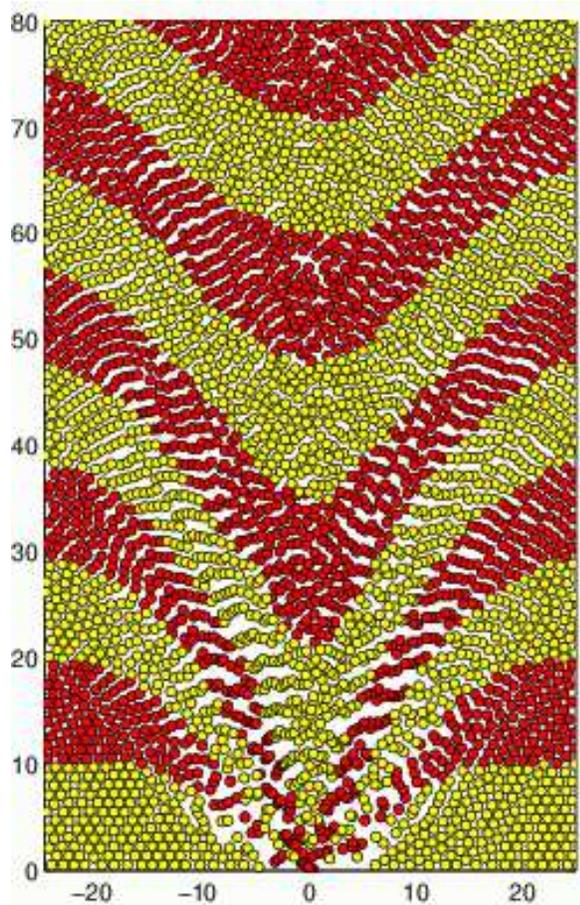}
\caption{ \label{fig:spot_layers} A  spot simulation of the experiment in
Fig.~\ref{fig:azadeh}, to be compared with the void  
simulation in Fig.~\ref{fig:void_layers}. 
(Courtesy of Jaehyuk Choi.)
}
\end{center}
\end{figure}

\begin{figure}
\begin{center}
\includegraphics[height=3.05in]{Figure4a_Exp_compress.ps} \nolinebreak
\includegraphics[height=3in]{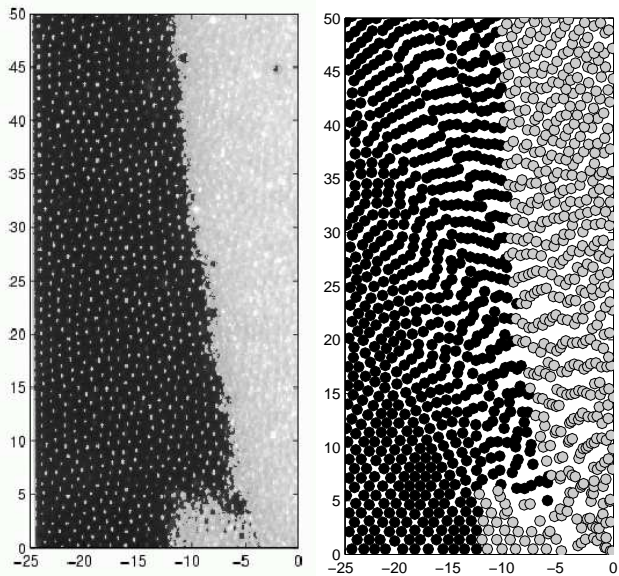}
\caption{ \label{fig:spot_interface} A spot-model simulation (right) of the
experiment in Fig.~\ref{fig:interface} (left). }
\end{center}
\end{figure}

For illustration purposes, we consider the simplest version of the
model in two dimensions without boundaries.  Let us assume that spots
undergo independent, directed random walks upward on a lattice
(without interactions). The vertical spot lattice parameter is set on
the order of a particle diameter, $d$, and the horizontal lattice
parameter is constrained by kinetic parameter, $b$, via
Eq.~(\ref{eq:bspot}).  Alternatively, for smoother spot motion using a
persistent random walk, the vertical lattice parameter may be reduced
for a desired level of auto-correlation, say $\gamma=0.5$, where $b$ is now
set by Eq.~(\ref{eq:bper}). We further assume
Eq.~(\ref{eq:spotsimple}) for the spot influence, where each particle
within a distance, $R$, from the spot center (after a displacement),
receives the same anti-parallel displacement, smaller by a factor, $w$.

Spot simulations of the experiments in Fig.~\ref{fig:azadeh} and
~\ref{fig:interface} are shown in Fig.~\ref{fig:spot_layers} and
~\ref{fig:spot_interface}, respectively. The spot parameters
determined from independent experimental measurements, as described
above ($b=1.3\, d$, $R=2.5\, d$, $w=0.0024$), without any {\it ad hoc}
fitting.  The result is a dramatic improvement over the simulations in
Figs.~\ref{fig:void_layers} and ~\ref{fig:interface}(c) of the same
situations using the Void Model, which produces far more
mixing. Remarkably, the mean flow profile is nearly the same in all
cases, given by Eq.~(\ref{eq:meanflow}), even though the particle
dynamics is radically different.

It is visually apparent that the Spot Model resolves the ``paradox of
granular diffusion''. Detailed analysis (to appear elsewhere) shows
that the experimental dynamics of two-color interfaces is reproduced
rather well by simulations, as in Fig.~\ref{fig:spot_interface}. The
interface remains sharp while moving with the mean flow and slowly
developing fluctuations at the scale of the spot size.

\subsection{ Density Fluctuations and Packing Constraints }

The reader may have noticed something strange about the simulations in
Figs.~\ref{fig:spot_layers} and ~\ref{fig:spot_interface} --- the
nearly uniform density of particles is not preserved.  In regions of
shear, the particles form dense overlapping bands separated by thin
vacuum bands transverse to the shear direction. Although mesoscopic
shear bands do form in disordered materials, such as metallic glasses
(see section~\ref{sec:outlook}), these bands are at the particle
scale, and hence are not physical. At the lower edges of the flow
region, where the shear rate is highest, the fluctuations intensify,
while directly above the orifice the density increases.

We conclude that the simplest version of the Spot Model does not
completely describe the joint probability distribution of all
particles, even though it predicts the marginal distribution of each
tracer particle quite well. As expected, what is missing is a strict
enforcement of packing constraints. The block-like cooperative motion
in spots causes density fluctuations to grow very slowly, but
unphysical structures eventually arise.

Stabilizing the density with a simple analytical modification is
likely to be difficult.  As discussed above, density-preserving extra
fluctuations appear to be related to the super-diffusive regime seen in
experiments, which has a nontrivial, sub-ballistic scaling. Moreover,
we should not expect to be able to generate a multitude of random
close packings by a simple statistical model. Even without flow, the
sampling of random close packings is a challenging computational
problem in its own right, which seems to require brute-force molecular
dynamics~\cite{torquato00,torquato01,torquato02,kansal02,ohern02,ohern03}.

On the other hand, unphysical density fluctuations also arise in the
standard kinetic theory of liquids, when particles undergo independent
random walks.  In dilute gases, this assumption correctly predicts
Poisson-distributed density fluctuations, where the variance in the
particle number scales with the volume~\cite{nicolas03}, but in a
liquid, just as in a granular material, this picture is
incorrect. Packing constraints during flow somehow preserve much
smaller, ``hyper-uniform'' density fluctuations, where the variance
scales with the surface area, as in the case of a regular
lattice~\cite{torquato03}. It is fair to say that a collective
statistical theory of liquids, molecular or granular, has not yet been
developed, so we should not be easily deterred.

\subsection{ Multiscale Algorithms for Dense Flows }

\begin{figure}
\begin{center}
\includegraphics[width=2.7in]{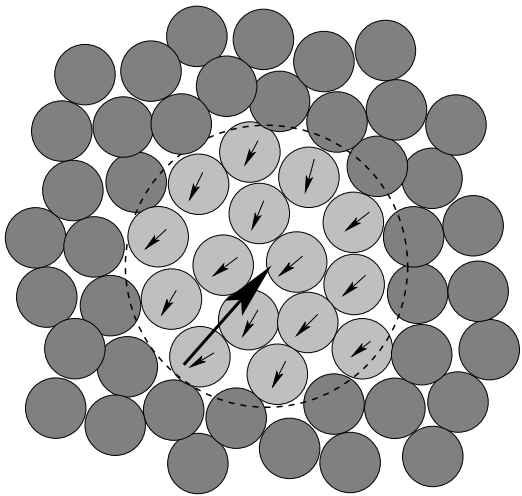}
\caption{ \label{fig:spot3} The Spot Hypothesis III: Real spots also
cause small internal rearrangements driven by packing constraints
(relative to the average motion shown in Fig.~\ref{fig:spot}).  In a
multiscale simulation algorithm, this could be done by a constrained
relaxation following the mean spot-induced displacements. }
\end{center}
\end{figure}

We have seen that the collective particle dynamics in the Spot Model
is quite realistic, aside from slowly growing density fluctuations, so
it may be that only occasional small perturbations could suffice to
preserve the density.  For example, after some (or all) spot-induced
cooperative displacements, the affected particles could be relaxed to
equilibrium with simple spring-like forces, perhaps including the
effect of gravity. The center of mass of the block of displaced
particles should be constrained during relaxation to prevent slipping
back into the initial positions.  A preliminary implementation of this
idea using a soft-core repulsion between grains (without gravity or
friction) produces rather realistic bulk flows with a stable density
~\cite{chris}.

Regardless of the detailed implementation, fixing the density will
result in some additional cooperative fluctuations relative to the
mean spot-induced displacement, as shown in Fig.~\ref{fig:spot3}. This
is surely a more realistic picture of a dense flow, which incorporates
the microscopic inter-particle forces into a new kind of multiscale
simulation algorithm. In spite of the potentially expensive relaxation
step, such an algorithm is much more efficient than brute-force
molecular dynamics tracking all particle contacts. It takes advantage
of spots to accelerate the dynamics, while keeping the system in a
nearly jammed state.

This would be an interesting direction to pursue, not only for
simulations, but also to gain insight into the basic physics of dense
flows. Unlike the mean spot-induced motion, the extra fluctuations to
preserve the density are correlated over the passage of multiple
spots. The result could be a crossover from super to normal diffusion
after a particle interacts with many spots, perhaps once it falls by
its own diameter.  Understanding such small-scale fluctuations is a
fundamental open question.

\section{ The Continuum Limit  }
\label{sec:math}

Macroscopic approximations offer many advantages over microscopic
theories, whenever a clear connection can be made. Both analytical
and numerical solutions are usually much easier starting from a
continuum approximation than an underlying discrete model. Moreover,
various poorly understood microscopic details are swept into
parameters in the continuum equations, which (one hopes) capture the
essential physics in a robust way. A good example is the simultaneous
description of Brownian motion, financial time series, and heat
transfer by the diffusion equation. Perhaps even more remarkable is
continuum hydrodynamics, which hides our ignorance of collective
statistical dynamics in liquids. In this section, we derive continuum
equations for diffusion and flow in the Spot Model based on the simple
picture of Fig.~\ref{fig:spot}, thus ignoring the complicated
density-preserving relaxation in Fig.~\ref{fig:spot3}.

\subsection{ A Nonlocal Stochastic Differential Equation }

\begin{figure}
\begin{center}
\includegraphics[width=\linewidth]{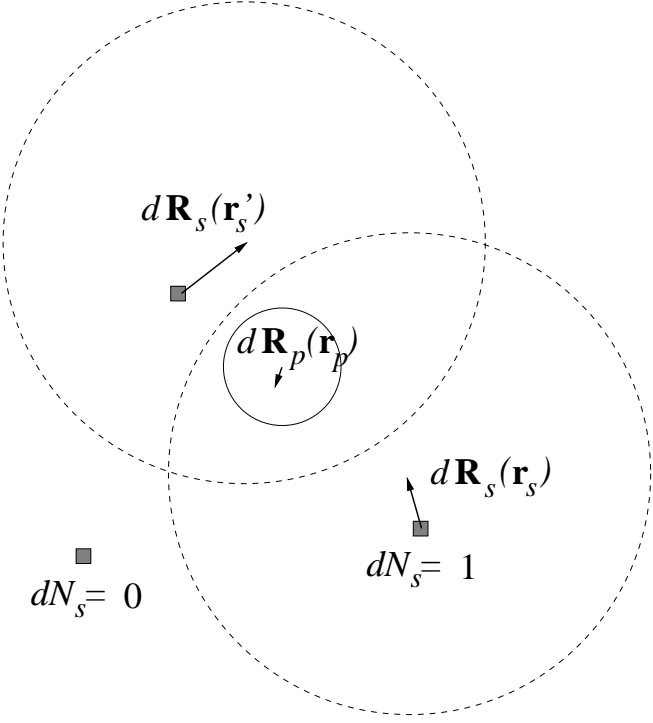}
\caption{ \label{fig:SDE_defs} Sketch of a particle interacting
with a collection of passing spots, showing some of the quantities
involved in defining the nonlocal SDE, Eq.~ (\protect\ref{eq:tracer}). }
\end{center}
\end{figure}

We begin by partitioning space as shown Fig.~\ref{fig:SDE_defs}, where
the $n$th volume element, $\Delta V_s^{(n)}$, centered at
$\rb_s^{(n)}$ contains a random number, $\Delta N_s^{(n)}$, of spots
at time $t$ (typically one or zero). In a time interval, $\Delta t$,
suppose that the $j$th spot in the $n$th volume element makes a random
displacement, $\Delta \Rb_s^{(j)}(\rb_s^{(n)})$ (which could be zero).
According to Eq.~(\ref{eq:discrete}), the total displacement, $\Delta
\Rb_p$, of a particle at $\rb_p$ in time $\Delta t$ is then given by a
sum of all the displacements induced by nearby spots,
\begin{equation} 
\Delta\Rb_p(\rb_p) = - \sum_n \,
\sum_{j=1}^{\Delta N_s^{(n)}}  w\left(\rb_p, \rb_s^{(n)} +
\Delta\Rb_s^{(j)}(\rb_s^{n})\right) \,
\Delta\Rb_j^{(n)}(\rb_s^{n})   \label{eq:tracer_discrete}
\end{equation} 
Note that the spatio-temporal distribution of spots,
$\Delta N_s^{(n)}$, is another source of randomness, in addition to
the individual spot displacements, $\Delta\Rb_j^{(n)}$, so that each
particle displacement is given by a random sum of random variables.

In the continuum limit, we arrive at a non-local, nonlinear stochastic
differential equation (SDE):
\begin{equation}
d\Rb_p(\rb_p,t) = - \int dN_s(\rb_s,t) \, w(\rb_p, \rb_s + d\Rb_s(\rb_s,t))
\, d\Rb_s(\rb_s,t) \label{eq:tracer} ,
\end{equation}
where the stochastic integral is defined by the limit of the random Riemann sum
in Eq.~(\ref{eq:tracer_discrete}).  This equation differs from
standard nonlinear SDEs in two basic ways: (i) The tracer trajectory,
\begin{equation}
\rb_p(t) = \int_{\tau=0}^t d\Rb_p(\rb_p,\tau)   \label{eq:trajectory}
\end{equation}
is passively driven by a stochastic distribution of moving influences
(spots), $dN_s(\rb_s,t)$, which evolves in time and space, rather than
by some internal source of independent fluctuations, and (ii)
the stochastic differential, $d\Rb_p(\rb_p,t)$, is given by a
non-local integral over other stochastic differentials, $d\Rb_s(\rb_s,t)$,
associated with these moving influences, which lie at finite
distances away from the particle.

\subsection{ Mean-field Fokker-Planck Equation  }

Equation~ (\ref{eq:tracer}) seems to be a new type of SDE, for which a
mathematical theory needs to be developed.  Here, we begin deriving
analytical results based on the assumption that all the stochastic
differentials for particle displacements in Eq.~(\ref{eq:trajectory})
are independent. This ignores the fact that the spot distribution at
one time, $dN_s(\rb_s,t)$, depends explicitly on the distribution and
displacements at the previous time, $t - dt$. Even for a stationary
spot distribution, it also ignores the fact that each spot, due to its
extended influence, generally affects the same particle more than
once, so that any autocorrelations in spot motion are partly
transferred to the particles. 

Assuming independent displacements, the propagator,
$P_p(\rb,t|\rb_0,t_0)$, which gives the probability density of finding
the particle at $\rb$ at time $t$ after being at $\rb_0$ at time
$t_0$, satisfies a Fokker-Planck (or forward Kolmogorov)
equation~\cite{risken},
\begin{equation}
\frac{\partial P_p}{\partial t} + \del\cdot(\ub_p P_p)
= \del\del : (\Db_p P_p)   \label{eq:fp} ,
\end{equation}
with drift velocity,
\begin{equation}
\ub_p(\rb,t) = \frac{\langle d\Rb_p(\rb,t)\rangle}{dt} = \lim_{\Delta
  t \rightarrow 0} \frac{\langle \Delta \Rb_p(\rb,t) \rangle}{\Delta
  t} ,
\end{equation}
and  diffusivity tensor, 
\begin{equation}
\Db_{p}^{\alpha\beta}(\rb,t) = \frac{\langle dR_p^{\alpha} 
dR_p^{\beta} \rangle}{2\,  dt} .
\end{equation}
(Here $\del\del : \Ab$ denotes $\sum_\alpha \sum_\beta
\frac{\partial^2 A^{\alpha\beta}}{\partial {x_\alpha} \partial
{x_\beta}}$.)  
The Fokker-Planck coefficients may be calculated by taking the
appropriate expectations using Eq.~(\ref{eq:tracer_discrete}) in the
limits $\Delta V_s^{(n)} \rightarrow 0$ and then $\Delta t \rightarrow
0$, which is straightforward if we assume (as above) that spots do not
interact. In this approximation, the spot displacements,
$\Delta\Rb_s^{(j)}(\rb_s^{(n)})$, and the local numbers of spots,
$\Delta N_s^{(n)}$, are independent random variables within each
infinitesimal time interval. They are also independent of the same
variables at earlier times, as assumed above.

In order to calculate the drift velocity, we need only the mean spot
density, $\rho_s(\rb_s,t)$, defined by $\langle \Delta N_s^{(n)}
\rangle = \rho_s(\rb_s,t) \, \Delta V_s^{(n)}$.  The result,
\begin{eqnarray}
\ub_p(\rb_p,t) &=& - \int dV_s \, w(\rb_p,\rb_s) \,
\left[ \rho_s(\rb_s,t) \ub_s(\rb_s,t) \right. \nonumber \\
& &  \left. - 2\, \Db_s(\rb_s,t) \cdot \del \rho_s(\rb_s,t) \right] \label{eq:up}
\end{eqnarray}
exhibits two sources of drift. The first term in the integrand simply
opposes the spot drift velocity,  
\begin{equation}
\ub_s(\rb,t) = \frac{\langle d\Rb_s(\rb,t)\rangle}{dt} ,
\end{equation}
as in the heuristic equation ~(\ref{eq:spotsimple}).
The second term, which depends on the spot diffusion tensor,
\begin{equation}
\Db_{s}^{(i,j)}(\rb,t) = \frac{\langle dR_s^{(i)} dR_s^{(j)}\rangle}{2\,
    dt} ,
\end{equation}
is a ``noise-induced drift'' (typical of nonlinear SDEs~\cite{risken})
causing particles to climb
gradients in the spot density. Both effects are
averaged over nearby regions, weighted by the spot influence function,
$w(\rb_p,\rb_s)$.

In order to calculate the diffusivity tensor, we also need information
about fluctuations in the spot density. Continuing with our
``mean-field approximation'', it is natural to assume independent
spot fluctuations, 
\begin{equation}
\langle \Delta N_s^{(n)} \Delta N_s^{(m)} \rangle =
\delta_{m,n} \langle (\Delta N_s^{(n)})^2\rangle = O\left((\Delta
V_s^{(n)})^\nu\right),
\end{equation}
where $\nu=1$ for a Poisson process and
$\nu<1$ for a hyper-uniform process~\cite{torquato03}.  It turns out 
that such fluctuations do not contribute to the diffusion
tensor (in more than one dimension), and the result is 
\begin{equation}
\Db_p(\rb_p,t) = \int dV_s \, w(\rb_p,\rb_s)^2 \rho_s(\rb_s,t) \,
\Db_s(\rb_s,t) .  \label{eq:Dp}
\end{equation}
Note that the influence function, $w$, appears squared in
Eq.~(\ref{eq:Dp}) and linearly in Eq.~(\ref{eq:up}). This causes the
P\'eclet number for tracer particles to be of order $1/w$ smaller than
that of spots (or free volume), as was derived heuristically above in
Eq.~(\ref{eq:bpsimple}).

An interesting general observation about Equation ~(\ref{eq:fp}) with
coefficients (\ref{eq:up}) and (\ref{eq:Dp}) is that rescaling the
spot density is equivalent to rescaling time. In other words, all
aspects of the stochastic motion of a tracer particle, including its
P\'eclet number, are determined by geometry, independent of the total
flow rate (or average spot density).  As explained above, this
experimental observation is in qualitative disagreement with kinetic
theory and quantitative disagreement with the Void Model.  On the
other hand, the fact that fluctuations are proportional to the spot
density suggests that a meaningful definition of ``temperature'' for
dense granular flows (if one exists) might simply depend on the free
volume of random packings. This possibility is discussed below in
section~\ref{sec:outlook}.

Higher-order terms a Kramers-Moyall expansion generalizing
Eq.~(\ref{eq:fp}) for finite independent displacements, which depend
on fluctuations in the spot density, are straightforward to calculate,
but beyond the scope of this paper.  Such terms are typically ignored
in stochastic analyses because, in spite of improving the overall
approximation, they tend to produce small negative probabilities in
the tails of distributions~\cite{risken}.  For dense granular flows
described by Eq.~(\ref{eq:tracer}), however, they may be important
since displacements can be relatively large compared to spatial
variations in probability density. This is the case in highly
nonuniform flows, such as near the orifice of a silo, where mean
velocities may vary on the scale of a few particle diameters.

\subsection{ Spatial Velocity Correlation Tensor }

For any stochastic process representing the motion of a single
particle, it is well-known that transport coefficients can be
expressed in terms of {\it temporal} correlation functions via the
Green-Kubo relations~\cite{risken}. For example, the diffusivity
tensor in a uniform flow is given by the time integral of the velocity
auto-correlation tensor,
\begin{equation}
D_p^{\alpha\beta} = \int_0^\infty dt \langle U_p^\alpha(t) U_p^\beta(0)
\rangle 
\end{equation}
where $\Ub_p(t) = \{ U_p^\alpha\} = d\Rb_p/dt$ is the stochastic
velocity of a particle. (A similar relation holds for spots.) 

In the Spot Model, nearby particles move cooperatively, so the
transport properties of the collective system also depend on the
two-point {\it spatial} velocity correlation tensor,
\begin{equation}
C_p^{\alpha\beta}(\rb_1,\rb_2) =
\frac{\langle U_p^\alpha(\rb_1) U_p^\beta(\rb_2) \rangle} 
{\sqrt{\langle U_p^\alpha(\rb_1)^2\rangle \langle
U_p^\beta(\rb_2)^2 \rangle }}   \label{eq:Cdef}
\end{equation}
which is normalized so that $C_p^{\alpha\beta}(\rb,\rb)=1$.  We
emphasize that the expectation above is conditional on finding two
particles at $\rb_1$ and $\rb_2$ at a given moment in time and
includes averaging over all possible spot distributions and
displacements.  Substituting the SDE (\ref{eq:tracer}) into
Eq.~(\ref{eq:Cdef}) yields
\begin{equation}
C_p^{\alpha\beta}(\rb_1,\rb_2) =
\frac{ \int dV_s \, \rho_s(\rb_s) \, D_s^{\alpha\beta}(\rb_s) \, 
w(\rb_1,\rb_s)\,  w(\rb_2,\rb_s) }
{\sqrt{ D_p^{\alpha\beta}(\rb_1)  D_p^{\alpha\beta}(\rb_2)}}  \label{eq:Cp}
\end{equation} 
where we have assumed again that spots do not interact (independent
displacements).  

Equation~(\ref{eq:Cp}) is a new kind of integral relation for
cooperative diffusion, which relates the spatial velocity correlation
tensor to the spot (or free volume) diffusivity tensor via integrals
of the spot influence function, $w(\rb_p,\rb_s)$.  If the statistical
dynamics of spots is homogeneous (in particular, if $\Db_s$ is
constant), then the relation simplifies:
\begin{equation}
C^{\alpha\beta}_p(\rb_1,\rb_2) = \frac{\int dV_s \, \rho_s(\rb_s) \,
w(\rb_1,\rb_s) \, w(\rb_2,\rb_s) }
{\sqrt{ \int dV_s \, \rho_s(\rb_s) \, w(\rb_1,\rb_s)^2 \,
 \int dV_s^\prime \, \rho_s(\rb_s^\prime) \, w(\rb_2,\rb_s^\prime)^2
}}.
\end{equation}
if $D_s^{\alpha\beta} \neq 0$ (and $0$ otherwise).  If the statistical
dynamics of particles is also homogeneous, as in a uniform flow
($\rho_s=$ constant), then it simplifies even further:
\begin{equation}
C^{\alpha\beta}_p(\rb) = \frac{\int dV_s \, w(\rb-\rb_s) 
\, w(-\rb_s) } { \int dV_s \, w(\rb_s)^2 }
\end{equation}
where we have assumed that the spot influence function, and thus the
correlation tensor, are translationally invariant ($\rb =
\rb_1-\rb_2$).  These predictions may be used to infer the spot
influence function experimental measurements, as in
Eq.~(\ref{eq:alpha}) above for uniform spots with sharp cutoff. For the case
of a symmetric Gaussian spot of width $\sigma$ in each direction,
\begin{equation}
w(\rb) = \frac{V_s}{(2\pi \sigma^2)^{3/2}} e^{-r^2/2\sigma^2} , \label{eq:gaussian}
\end{equation}
an even simpler formula results from the overlap integral,
\begin{equation}
C_p^{\alpha\beta}(\rb) = e^{-r^2/4\sigma^2}.
\end{equation}

\subsection{ Relative Diffusion of Two Tracers }

The spatial velocity correlation function affects many-body transport
properties. For example, the relative displacement of two tracer
particles, $\rb = \rb_1 - \rb_2$, has an associated diffusivity
tensor given by,
\begin{eqnarray}
D^{\alpha\beta}(\rb_1,\rb_2)& =& D_p^{\alpha\beta}(\rb_1) +
D_p^{\alpha\beta}(\rb_2) \\
& &  - 2 \, C^{\alpha\beta}_p(\rb_1,\rb_2)\,
\sqrt{ D_p^{\alpha\alpha}(\rb_1)  D_p^{\beta\beta}(\rb_2) }  \nonumber
\end{eqnarray}
In  a uniform flow, the diagonal components take the simple form
\begin{equation}
D^{\alpha\alpha}(\rb) = 2\, D^{\alpha\alpha}_p \left( 1 -
C^{\alpha\alpha}_p(\rb)\right) 
\end{equation}
which was used above to estimate the cage-breaking length. A more
detailed calculation of the relative propagator, $P(\rb,t|\rb_0,t_0)$,
neglecting temporal correlations (as above) 
would start from the associated Fokker-Planck equation,
\begin{equation}
\frac{\partial P(\rb,t)}{\partial t} = \del\del : \left(\Db(\rb)
P(\rb,t)\right)  \label{eq:fprel}
\end{equation}
with a delta-function initial condition. (In a non-uniform flow, one
must also account for spurious drift and motion of the the center of
mass.)

Our simplified analysis does not enforce packing constraints, so it
allows for two particles to be separated by less than one diameter. A
hard-sphere repulsion may be approximated by a reflecting boundary
condition at $|\rb|=d$ when solving equations such as
(\ref{eq:fprel}), but there does not seem to be any simple way to
enforce inter-particle forces exactly. However, this may be precisely
why the model remains mathematically tractable and intuitive, while
capturing the essential physics of diffusion in dense flows.

\section{ Application to Silo Drainage }
\label{sec:spotcont}

\subsection{ Statistical Dynamics of Spots }

The analysis in the previous section makes no assumptions
about spots, other than the existence of well-defined local mean
density, mean velocity, and diffusion tensor, which may depend on time
and space.  As such, the results may have relevance for a variety of
dense disordered systems exhibiting cooperative diffusion (see
below). In this section, we apply the general theory to the specific case of
granular drainage, in which spots diffuse upward from a silo orifice.

For simplicity, let us assume that each spot undergoes mathematical Brownian
motion with a vertical drift velocity, $\ub_s = v_s \hat{z}$,
and a diagonal diffusion tensor,
\begin{equation}
\Db_s = \left( \begin{array}{ccc}
D_s^\perp & 0 & 0 \\
0 & D_s^\perp  & 0 \\
0 & 0 & D_s^\|
\end{array} \right)   \label{eq:Ds}
\end{equation}
which allows for a different diffusivity in the horizontal ($\perp$)
and vertical ($\|$) directions due to the symmetry-breaking effect
of gravity. In that case, the propagator for a single
``spot tracer'',
$P_s(\xb,z,t|\xb_0,z_0,t)$, satisfies another Fokker-Planck equation,
\begin{equation}
\frac{\partial P_s}{\partial t} + \frac{\partial }{\partial
z}\left(v_s P_s\right) =
\del^2_\perp \left( D_s^\perp P_s \right) +
 \frac{\partial^2 }{\partial z^2}\left(D_s^\| P_s\right) .
 \label{eq:spotfp} 
\end{equation} 
The coefficients may depend on space (e.g. larger velocity above the
orifice than near the stagnant region), 
as suggested by the experimental density-wave measurements 
discussed above~\cite{behringer89}. 

The geometrical spot propagator, $\Pc_s(\xb,|z,\xb_0,z_0)$, is the
conditional probability of finding a spot at horizontal position $\xb$
once it has risen to a height $z$ from an initial position
$(\xb_0,z_0)$. For constant $v_s$ and $\Db_s$, the geometrical
propagator satisfies the diffusion equation, 
\begin{equation}
\frac{\partial \Pc_s}{\partial z} = b\, \del^2_\perp \Pc_s \label{eq:spotdiff}
\end{equation} 
where $b = D_s^\perp v_s$ is the kinematic parameter.  If spots move
independently, this equation is also satisfied by the steady-state
mean spot density, $\rho_s(\xb,z)$, analogous to Eq.~(\ref{eq:rhov})
for the void density in the Void Model. However, the mean particle
velocity in the Spot Model, Eq.~(\ref{eq:up}), is somewhat different,
as it involves nonlocal effects (see below).

The time-dependent mean density of spots, $\rho_s(\xb,z,t)$, depends
on the mean spot injection rate, $Q(\xb_0,z_0,t)$
(number/area$\times$time), which may vary in time and space due to
complicated effects such as arching and jamming near the orifice. It
is natural to assume that spots are injected at random points along
the orifice (where they fit) according to a space-time Poisson process
with mean rate, $Q$. In that case, if spots do not interact, the
spatial distribution of spots within the silo at time $t$ is also a
Poisson process with mean density,
\begin{eqnarray}
\rho_s(\xb,z,t) &=& \int d\xb_0 \int dz_0 \int_{t_0<t} dt_0 \nonumber \\
& & Q(\xb_0,z_0,t)\, P_s(\xb,z,t|\xb_0,z_0,t_0) . \label{eq:rhos}
\end{eqnarray}
For a point-source of spots (i.e. an orifice roughly one spot wide)
at the origin with flow rate, $Q_0(t)$ (number/time), this reduces to
\begin{equation}
\rho_s(\xb,z,t) = \int_{t_0<t} dt_0 \, Q_0(t_0)\, P_s(\xb,z,t|0,0,t_0) ,
\end{equation}
where $P_s$ is the usual Gaussian propagator for Eq.~(\ref{eq:spotfp})
in the case of constant $u_s$ and $\Db_s$. In reality, spots should
weakly interact, but the success of the Kinematic Model suggests that
spots diffuse independently as a good first approximation.

\subsection{ Statistical Dynamics of Particles }

Integral formulae for the drift velocity and diffusivity tensor of a
tracer particle may be obtained by substituting the spot density
which solves Eq.~(\ref{eq:rhos}) into the general expressions ~(\ref{eq:up}) and
(\ref{eq:Dp}), respectively.  For example, if spots only diffuse
horizontally ($D_s^\|=0$), then the mean downward velocity of
particles is given by 
\begin{equation}
v_p(\rb,t) =  \int dV_s \, w(\rb_p,\rb_s)\, \rho_s(\rb_s,t) \, v_s(\rb_s,t)
\end{equation}
The main difference with the analogous prediction of the Void Model, Eq.~
(\ref{eq:voidflux}), is the convolution with the spot influence
function. These sorts of integrals are most important in regions of
high shear, where the spot density and/or spot dynamics varies on
scales comparable the spot size (several particles wide). 

For simplicity, let us consider a bulk region where the
spot density varies on scales much larger than the spot size. In this
limit, the integrals over the spot influence function reduce to the
following ``interaction volumes'':
\begin{equation}
V_k(\rb) = \int d\rb_s w(\rb,\rb_s)^k
\end{equation}
for $k=1,2$. (Note that $V_1=V_s$ above.) The equation
for tracer-particle dynamics (\ref{eq:fp}) then takes the form,
\begin{eqnarray} 
\frac{\partial P_p}{\partial t} &=& \frac{\partial }{\partial
z}\left[\left( v_s \rho_s - 2 D_s^\| \frac{\partial \rho_s}{\partial z}
\right) V_1 P_p \right] \nonumber \\
& &  - 2 \del_\perp\cdot\left( D_s^\perp (\del_\perp \rho_s)
V_1 P_p\right)  
\nonumber \\
& & 
 \frac{\partial^2 }{\partial z^2}\left(D_s^\| \rho_s V_2 P_p \right)
+
\del^2_\perp \left( D_s^\perp \rho_s V_2 P_p \right) .
 \label{eq:spotdrainfp} 
\end{eqnarray} 
Again, it is clear that rescaling the spot density is equivalent to
rescaling time.

When the spot dynamics is homogeneous (i.e. $u_s$ and $\Db_s$ are
constants), Equation~(\ref{eq:spotdrainfp}) simplifies further:
\begin{eqnarray}
\frac{1}{v_s V_s} 
\frac{\partial P_p}{\partial t} &=& \left( \frac{\partial }{\partial
z} + b_p^{\perp} \del^2 + b_p^{\|}
 \frac{\partial^2 }{\partial z^2}\right)(\rho_s P_p)  \label{eq:simplespotfp} 
\\
& &  - 2 b^\perp \del\cdot(P_p \del \rho_s) 
- 2 b^\| \frac{\partial}{\partial z}\left(P_p \frac{\partial
\rho_s}{\partial z}\right)  \nonumber 
\end{eqnarray} 
where $b^\perp = b = D_s^\perp/v_s$ and $b^\| = D_s^\|/v_s$ are the
spot diffusion lengths and $b_p^\perp = b_p V_2/V_1$ and $b_p^{\|} =
b^{\|} V_2/V_1$ are the particle diffusion lengths. In this
approximation, the latter are given by the simple formula,
\begin{equation}
\frac{b_p^{\perp}}{b^\perp} = \frac{b_p^\|}{b^\|} 
= \frac{\int dV_s \, w(\rb,\rb_s)^2 }{\int dV_s w(\rb,\rb_s)}  \label{eq:bp}
\end{equation}
which generalizes Eqs.~(\ref{eq:bpsimple}) and (\ref{eq:bw}) derived
above for a uniform spot with a sharp
cutoff. In the case of a Gaussian spot, Eq.~(\ref{eq:gaussian}), we
obtain a simple relation,
\begin{equation}
\frac{b_p^{\perp}}{b^\perp} = \frac{b_p^\|}{b^\|} =
\frac{V_s}{8\pi^{3/2}\sigma^3}
\end{equation}
between the particle and spot diffusion lengths and the spot size and
free volume.

The physical meaning of the diffusion lengths becomes more clear in
the limit of uniform flow, $\rho_s =$ constant. In terms of the
position in a frame moving with the mean flow, $\zeta = v_p t - z$,
where $v_p = v_s V_s \rho_s$, we arrive at a simple diffusion
equation,
\begin{equation}
\frac{\partial P_p}{\partial \zeta} = \left( b_p^{\perp} \del_\perp^2 +
b_p^{\|} \frac{\partial^2}{\partial z^2}\right) P_p ,
\end{equation}
where $\zeta$ acts like time. Therefore, $b_p^{\perp}$ and $b_p^{\|}$
are the variances of the displacements in the perpendicular and
parallel directions per twice the mean distance dropped.

Solutions to the new equations above and comparisons with
experimental data are left for future work.

\section{ Outlook for Granular Materials }
\label{sec:outlook}

\subsection{ Silo Drainage as a Physics Problem }

Gravity-driven silo drainage is an ideal system for fundamental
studies of diffusion and mixing.  The simplicity of the setup, free of
externally applied shear, rotation, or vibration, may make it the
easiest to understand in microscopic detail.  Future work should focus
on detailed comparisons between the present theory (both discrete
models and continuum equations), experiments, and granular dynamics
simulations. It is natural to start with dry, cohesionless,
monodisperse grains, as in this work, and then move on to more
complicated granular materials.  Surely many more physical principles
remain to be discovered.  An important goal should be connect with an
appropriate continuum model, perhaps among those cited in the
introduction.  Few of these models have been carefully tested in
silo-drainage experiments, and none has been derived systematically
from a microscopic theory including diffusion.

One such model, due to Aranson {\it et al.}~\cite{aranson99,aranson01}
postulates the co-existence of ``liquid-like'' and ``solid-like''
micro phases in the granular flow, where the local liquid fraction is
an order parameter following {\it ad hoc} Landau-type dynamics. (This
is a substantial generalization of models of surface avalanches, which
postulate co-existing static and rolling
phases~\cite{bouchaud94,boutreux98}, and bears some similarity to
models of the glass transition~\cite{cohen59,cohen79}.)  The length
scale for order-parameter variations is assumed to be of order the
particle diameter, even though this might seem too small for a
continuum theory of co-existing phases.

In spite of its successes in describing experimental flows, the order
parameter lacks any clear microscopic basis. The present work,
however, suggests that it may be related to the spot density, at least
in the case of silo drainage (which has not yet been analyzed). If a
successful connection could be made with statistical dynamics, it
would be reminiscent of the microscopic theory of superconductivity,
which came decades after the phenomenological macroscopic description
of Ginzburg and Landau~\cite{aranson01}. Such connections may also
exist with other empirical continuum theories of granular and glassy
systems, which invoke the ``free volume per atom'' as a sort of order
parameter controlling material response.

\subsection{ Some Open Questions about Spots }

There are many questions about silo drainage raised by this work, once
one accepts the existence of spots.  {\it How do spots move?} Does
their drift velocity, diffusivity, and/or shape depend on particle
properties, position in the silo, or the presence of other spots? How
do spots interact with container walls or free surfaces?  {\it How do
particles move within spots?}  What are the ``extra'' fluctuations
associated with small-scale super-diffusion and the strict enforcement
of packing constraints?  Does the spot influence depend on the local
velocity, shear rate, or stress, the friction and elasticity of the
particles, etc.? {\it How do spots interact?} What kind of attraction
and repulsion between spots is responsible for stabilizing the volume
fraction in the range of flowing random packings? Do spots subdivide
and recombine? {\it How are spots created and destroyed?} How do spots
behave at an orifice or free surface? How do dilation and compaction
create and destroy spots in the bulk?

Even more basic questions have to do with the foundations of the
theory. {\it Why do spots even exist?} In a general sense, cooperative
motion arises from strong dissipation in the presence of packing
constraints (as in gravitational inelastic
collapse~\cite{kadanoff99,ertas02}), but why should the motion be
driven by coherent spots of free volume as described here? This
motivates the question of whether spot properties can be predicted
from first principles. If so, what are the key ingredients of a
complete {\it ab initio} theory? It seems that spots are mainly a
consequence of geometrical constraints in flowing, dense random
packings, but this connection should be made more precise. The
statistical geometry of random packings~\cite{torquato} may be an 
appropriate starting point.

\subsection{ Other Gravity Driven Flows }

Since there is strong evidence for spots in silo drainage, it is
natural to assume they also exist in other gravity-driven flows. In
the case of flow down an inclined chute~\cite{pouliquen99},
cooperative motion has also been invoked, albeit without a precise
microscopic description. Erta{\c s} and Halsey have explained the
Bagnold scaling of the mean velocity in terms of ``granular eddies'':
coherent structures, several particles in diameter, which rotate
rigidly in the same sense (as if rolling down the
plane)~\cite{ertas02}.  However, no evidence for granular eddies has
yet been found in experiments or granular dynamics
simulations~\cite{landry_note}.

The basic argument for granular eddies~\cite{ertas02} could also be
made for spots: Strong dissipation (``gravitational
collapse''~\cite{kadanoff99}) causes particles to move cooperatively
with their nearest neighbors. Perhaps some composite picture is
appropriate for chute flows, although internal rotation with the shear
may not be necessary.  With simple spots, shear can occur, as in the
flow region in a draining silo, when there is a gradient in the spot
density transverse to the mean spot drift.  With a theory of how spots
interact with the free surface, it may be possible to explain the data
without including internal same-sense rotations (or eddies) in the
Spot Model.

In the context of similar experiments in an inclined
pipe~\cite{pouliquen96}, Pouliquen {\it et al.} have proposed a
totally different mechanism for dense shear
flow~\cite{pouliquen01}. They postulate that flow occurs by a nonlocal
fluctuation-activated process: A shear event in one location initiates
stress fluctuations which can trigger or enhance shear in another
location by exceeding a local Coulomb yield criterion.  A simple
probabilistic model of this process describes the experimental mean
flow across the pipe cross section fairly well, but the microscopic
picture based on single-particle hopping into ``holes'' seems dubious,
in light of the present work. (``A particle... will jump to the next
hole with a probability equal to the probability that the stress
fluctuation is higher than the threshold''~\cite{pouliquen01}.)  As in
the Void Model for drainage, this picture should not be taken
literally since cooperative motion is likely to occur. Perhaps,
instead, a spot is triggered to move in the opposite direction by
stress fluctuations in response to other spot displacements. Such
interactions are not considered in the present model and might be
important in more general situations.

\subsection{ Forced Shear Flows }

Shear flows can also be driven by moving rough surfaces in granular
Couette cells~\cite{mueth00}. Such force shear flows seem quite
different from the gravity-driven flows discussed above, but some
similarities arise, which could be signs of the sort of cooperative
diffusion described here. It is usually found that the shear usually
localizes in a narrow band at the inner cylinder, as in some pipe
flows, but recently broad shear bands have also been observed by
Fenistein and van Hecke ~\cite{fenistein03}. If outer part of the
lower wall rotates with the outer cylinder while the inner part
remains fixed with the inner cylinder, then a region of shear smoothly
connects stagnant regions on either side of the interface. Rising
through the Couette cell, the shear region broadens and then drifts
toward the inner cylinder where it eventually localizes.

The situation is reminiscent of the broad flow region in silo
drainage, aside from the direction of the velocity (in the horizontal
plane of rotation).  In both cases, the velocity profile has a
self-similar form reminiscent of diffusion problems. In the Couette
flow, it is an error function, as if connecting two uniform
concentrations, rather than a Gaussian (for a point source). Unlike
gravity-driven drainage, the scaling of the shear zone is not that of
simple diffusion, since its width grows somewhat faster than the
square root of height, prior to localization near the inner cylinder.
Nevertheless, it may be that some kind of spot-like cooperative
diffusion originating with dilation at the shearing interface and
propagating upward may explain the velocity profile of the shear band.

\subsection{ Statistical Thermodynamics of Dense Flows? }

As discussed in the introduction, classical statistical thermodynamics
does not apply to slow granular flows, because the usual definition
of temperature, in terms of undissipated internal energy, makes no
sense. In the opposite limit of static granular systems, some
radically different definitions have been proposed. For example,
Ngan's ``mechanical temperature'' attempts to describe frozen
configurational entropy in terms of contact-force
distributions~\cite{ngan03}.

Edwards' thermodynamic theory of granular
compaction~\cite{edwards01,edwards91,edwards94,edwards98} focuses
instead on excess volume and is thus closer in spirit to the Spot
Model. His basic assumption is that all jammed configurations at a
given volume are equally probable. This form of ergodicity is
analogous to Boltzmann's famous postulate that all states at a given
energy are equally probable.  

Following this analogy, Edwards defines the ``compactivity'',
analogous to temperature, as the derivative of volume with respect to
entropy (number of jammed configurations).  The compactivity can also
be related to volume fluctuations, e.g. as measured in vibrational
experiments~\cite{nowak98}, just as the usual temperature is related
to velocity fluctuations.  Recent granular simulations have
demonstrated an effective Einstein relation in bulk shear flows, where
the ``temperature'', defined as the ratio of the diffusivity to the
mobility (in response to a force pulling on one grain), is consistent
with the Edwards compactivity~\cite{makse02}. This suggests that the
Edwards ensemble may have relevance for dense flows, although it is
not clear how to predict the diffusivity or derive equations for
hydrodynamics.

It is tempting to speculate that the spot density in silo drainage is
somehow related to the Edwards temperature (or compactivity).  Indeed,
the diffusivity is particles is proportional to the spot density, so
the Spot Model distinguishes between a ``warm'' region of fast
dynamics and ``cold'' region of slow dynamics by the amount of free
volume. This also seems consistent with viewing the compactivity as an
intensive variable controlling diffusion.

The drift of particles, however, is also proportional to the spot
density, and thus, if the analogy were correct, to temperature. This  
troubling implication, which violates the notion of ``temperature'' as
a measure of fluctuations, is related to the experimental fact that
fluctuations depend on the distance dropped, not
time~\cite{choi04}. The reader may conclude that any attempt at a
thermodynamic description of dense granular flows is unlikely to
succeed.

On the other hand, the Spot Model may provide a basis for some kind of
thermodynamics.  Rather than describing the system of strongly
interacting particles, which is wrought with difficulties related to
geometrical packing constraints, perhaps we should consider a system
of weakly interacting spots. We have already seen that this
gives a simple explanation of density waves, but perhaps it could also
be used to recover thermodynamics.  The spot diffusivity would be
determined by a ``spot temperature'' (perhaps related to the Edwards
definition) and the spot drift by independent external forcing due to
gravity.  Even if spots were to obey classical statistical mechanics
(starting from a ``spot Hamiltonian''), the particles would behave rather
differently,  with drift strongly coupled to diffusion.

\subsection{ Compaction Dynamics }

Even in the absence of flow, granular materials exhibit nontrivial
relaxation phenomena, such as compaction (or densification) under
vibration.  Experiments on vertically vibrated granular materials have
revealed a very slow $(\log t)^{-1}$ decay of
free volume with time~\cite{knight95,nowak98}.  To explain this
result, Boutreux and de Gennes~\cite{boutreux97} borrow  classical
equations from glass theory, due to Cohen and Turnbull~\cite{cohen59} (see
below): (i) They assume a Poisson distribution of ``holes'' of volume
$V$,
\begin{equation}
p(V) = \frac{1}{V_f}e^{-V/V_f}   \label{eq:Poisson}
\end{equation}
where $V_f$ is the local ``free volume'', defined (albeit
ambiguously~\cite{torquato,torquato00,kansal02}) so that $V_f=0$ in
the hypothetical ``random close packing''. (ii) The empirical
relation, $V_f \propto (\log t)^{-1}$, then follows from the
mean-field assumption that the rate of compaction, $dV_f/dt$, is
proportional to the probability
\begin{equation}
P(V_0) = e^{-V_0/V_f}   \label{eq:freevol}
\end{equation}
of finding a hole larger than a particle volume, $V_0$. Such a hole,
capable of being filled by a single particle, corresponds to a
``void''~\cite{nowak98}.

This thinking reflects the entrenched notion that structural
rearrangement in an amorphous material occurs by the independent
displacement of a particle into a randomly appearing void. Unlike
hypothetical voids in granular drainage, which propagate by exchanging
places with particles, those in granular compaction are annihilated by
hopping particles.  This picture of compaction has been compared
to the ``parking-lot model'' of cars trying to fill a dwindling set of
available spaces~\cite{nowak98}.

We have shown that voids do not play a significant role in dense
granular drainage, so it seems unlikely that they could also occur in 
compaction dynamics, which occurs at ever higher volume fractions,
closer to the jammed state. We would also expect nearest-neighbor
cages to remain largely preserved during compaction, especially in the
later stages, when the density is much closer to the jammed state than
in the drainage flows considered here (which exhibit very slow cage
breaking). It seems more plausible that particles rearrange in a
cooperative fashion during compaction, slowly eliminating interstitial
free volume without ever opening any particle-sized voids.

The Spot Model provides a general theoretical framework to describe
such collective mechanisms for transporting free volume from the bulk
to the upper surface.  Note that a new microscopic mechanism need not
contradict the prior successes of mean-field free-volume models.  As
with the mean velocity of granular drainage, the mean compaction rate
could be very similar with spots and voids, even though diffusion and
cage-breaking are completely different in the two cases. The
characteristic free volume, $v_0$, in Eq.~(\ref{eq:freevol}) should be
related to the typical volume carried by a spot, which is roughly a
tenth of a particle volume in granular drainage. This is consistent
with the more general view of $v_0$ as the ``typical free-volume
fluctuation required for a local collapse'' (see
below)~\cite{lemaitre02}.

\section{ Outlook for Glasses }
\label{sec:glassy}

\subsection{ Glassy Relaxation}

A major reason for the current excitement about granular materials in
physics is the analogy with glasses, which exhibit similar features
such as ``structural temperatures''~\cite{kurchan01} and a ``jamming
transition''~\cite{ohern03}. The hope is that such common features
could shed light on the nature of glass and the glass
transition~\cite{angell00}, which has been called ``the deepest and
most interesting unsolved problem in solid state
theory''~\cite{anderson95}. As in the case of granular flow, many toy
models and continuum approaches have been proposed, but glassy
relaxation remains poorly understood at the microscopic level.

It has long been recognized that cooperative motion plays a crucial
role in glassy relaxation, although precise mechanisms remain elusive.
Early theories of the diverging relaxation time at the
glass-transition temperature~\cite{vogel21,williams55} were based on
the hypothesis of free-volume
diffusion~\cite{eyring36,fox50,bueche53,cohen59}.  Adam and Gibbs
first suggested instead that free volume is associated with
temperature-dependent regions of cooperative relaxation, whose size
diverges at the glass transition~\cite{adam65}.  Cohen and Grest later
introduced more explicit correlations into free-volume theory (see
below) by considering clusters of ``liquid-like cells'', so the
melting of glass could be understood as a dynamical percolation
transition~\cite{cohen79}. The mode-coupling theory of G\"otze and
Sj\"ogren has invoked the ``cage effect'' of local blocking at the
scale of single particles to explain the emergence of slow
dynamics~\cite{gotze96}, albeit without describing how it occurs in
real space. In contrast, Berthier has provided new simulation evidence
for the Adam-Gibbs hypothesis that the glass transition is associated
with dynamical correlations at a length scale which grows with
decreasing temperature~\cite{berthier04}.

Experiments and simulations have revealed ample signs of ``dynamical
heterogeneity'' in supercooled liquids and spin glasses~\cite{hetero},
but the direct observation of cooperative motion has been achieved
only recently.  Rather than compact regions of local arrangement,
Donati {\it et al.} have observed ``string-like'' relaxation in
molecular dynamics simulations of a Lennard-Jones model
glass~\cite{donati98}. The strength and length scale of correlations
increases with decreasing temperature~\cite{donati99}, consistent with
the Adam-Gibbs hypothesis. These observations have motivated Garrahan
and Chandler to propose a phenomenological lattice model where each
site possesses an arrow pointing in the ``mean
direction of facilitation'' (opposite to the preferred direction of
flow) and chain-like relaxation follows from dynamical interactions
between the arrows ~\cite{garrahan03}. (The idea of a local vector
state variable also arises in theories of plasticity, discussed
below.)

Cooperative motion would be difficult to observe experimentally in a
molecular glass, but Weeks {\it et al.} have used confocal microscopy
to reveal three-dimensional clusters of fast-moving particles in dense
colloids~\cite{weeks00}. In the supercooled liquid, clusters of
cooperative relaxation have widely varying sizes, which grow as the
glass transition is approached. In the glass phase, the clusters are
much smaller, on the order of ten particles, and do not produce
significant rearrangements on experimental time scales. 

These observations suggest that spots may have relevance for
cooperative diffusion in simple glasses.  String-like relaxation is
reminiscent of the trail of a spot (see
Fig.~\ref{fig:rising_spots}). Atomic chains would correspond to the
motion of spots roughly one particle in size and carrying less than
one particle of free volume. Larger regions of correlated motion might
involve larger spots and/or collections of interacting spots. Key
features of the experimental data~\cite{weeks00} seem to support this
idea: (i) Correlations take the form of ``neighboring particles moving
in parallel directions'' (as in Fig.~\ref{fig:spot}); and (ii) the
large clusters of correlated motion tend to be fractals with dimension
two, as would be expected for the random-walk trail of a spot (as in
Fig.~\ref{fig:rising_spots} without the bias of gravity). It seems
more experiments and simulations should be done to test the spot
hypothesis for various glass-forming systems.

\subsection{ Free-Volume Theory }

For metallic glasses~\cite{johnson99,busch00}, free-volume theory
provides the basis for modern continuum models of plastic flow and
diffusion. The concept of hole diffusion was apparently introduced in
1936 by Eyring~\cite{eyring36}, but Cohen and Turnbull in 1959
initiated the standard model of glassy
relaxation~\cite{cohen59,turnbull61,turnbull70}, based on
Eqs.~(\ref{eq:Poisson}) and (\ref{eq:freevol}). In their theory of
``molecular transport in liquids and glasses'' (on the same footing),
randomly distributed ``holes'' are filled by single-particle
displacements, whenever a hole is large enough to accommodate a
neighboring particle, with probability $P(v_0) = e^{v_0/v_f}$. These
assumptions yield the classical formula, $D \propto e^{-v_0/v_f}$, for
the diffusivity, where a linear dependence of free volume on
temperature, $v_f = A + B(T-T_g)$, above the glass transition, $T_g$,
is commonly assumed, as suggested by Williams, Landel, and
Ferry~\cite{williams55}. By assuming that hole diffusion is biased by
the local shear stress, Spaepen derived Doolittle's
formula for the viscosity, $\eta \propto
e^{v_0/v_f}$, and proposed mechanisms for the creation and
annihilation of free volume based on particles jumping into holes of a
smaller or larger volume, respectively~\cite{spaepen77}.

Of course, this microscopic picture is very similar to that of the
Void Model for granular drainage, where gravity provides a bias for
void diffusion (so voids are created at the orifice and destroyed at
the free surface).  Indeed, Spaepen's Figure 3 ~\cite{spaepen77} is
nearly identical to Figure~\ref{fig:void} above.  The key point is
that, in both cases, flow and diffusion occur when a single particle
moves into a void independently from its neighbors.

Since cooperative diffusion is widely believed to occur in glasses,
the Spot Model seems to provide a more realistic mechanism for free
volume diffusion. As we have demonstrated for granular drainage, this
could be checked in glasses by comparing the diffusion lengths for
free volume and particles and the cage-breaking length (which must all
be roughly the same for hole diffusion) or by directly observing
spatial velocity correlations. If interacting spots exist in metallic
glasses, perhaps they are responsible for the formation of shear bands
of reduced density under large tensile loading~\cite{rajesh}.  Shear
bands in metallic glasses are difficult to understand, because, unlike
shear bands in crystals, which are known to result from dislocation
interactions, it is not clear what are the microscopic carriers of
plastic deformation in disordered materials.

\subsection{ Plastic Deformation }

The idea of localized cooperative motion has a long history in the
plastic deformation of amorphous solids.  Orowan~\cite{orowan52} was
perhaps the first to propose, contrary to Erying's hole-diffusion
hypothesis, that plastic flow occurs via localized shear
transformations in regions of enhanced atomic disorder (and free
volume)~\cite{orowan52}.  In the context of metallic glasses,
Argon~\cite{argon79} later distinguished between two types of local
rearrangements: (i) a ``diffuse shear transformation'' at high
temperature, in which a spherical region 4-6 atoms in diameter deforms
smoothly into an ellipsoidal shape, and (ii) an ``intense shear
transformation'' at low temperature, in which an equally small
disk-shaped region, similar to a dislocation loop, slides suddenly
along an atomic slip plane. A generalization of this concept, the
``localized inelastic transformation'' (LIT), forms the basis of a
stochastic model of elasto-plastic deformation due to Bulatov and
Argon~\cite{bulatov94}.  Each LIT acts irreversibly on a material
element as a pure shear along some discrete axis of symmetry plus an
additional dilation.

The basic idea of the LIT has recently been revived in
physics~\cite{STZnote} as a ``shear transformation zone''
(STZ)~\cite{falk98}, connecting ``$+$'' and ``$-$'' states relative to
the direction of an applied shear.  Falk and Langer first identified
STZ events in molecular dynamics simulations of a Lennard-Jones glass
and proposed a new continuum theory of viscoplastic deformation, which
involves the densities, $n_\pm$, of the two hypothetical STZ types as
auxiliary continuum variables (or order
parameters)~\cite{falk98}. They noted that STZ transition rates should
depend on the local free volume, taken to be constant as a first
approximation.  STZ theory compares favorably with experimental data
on metallic glasses~\cite{falk04}, although it contains a number of
adjustable parameters and {\it ad hoc} functional forms.

Lema\^itre has recently extended the two-state STZ theory to account
for the creation and annihilation (but not diffusion) of free
volume~\cite{lemaitre02}, borrowing some equations from the literature on
metallic glasses and granular compaction discussed above. The
resulting empirical constitutive laws are meant to apply to a broad
class of sheared dense materials, including glasses~\cite{lemaitre02b}
and granular materials~\cite{lemaitre02c}. The same phenomenology
provides a fairly simple way of understanding universal features of
amorphous systems, such as stick-slip, dilatency, and nonlinear
rheology, although again the microscopic mechanisms remain
vague. Moreover, if free volume is created and destroyed by particle
displacements, it seems it should also diffuse by very similar
mechanisms (as in Spaepen's theory~\cite{spaepen77}).

Although neglecting diffusion, Lema\^itre offers a new, qualitative
interpretation of free-volume kinetics, of particular interest here:
``Borrowing on the argument of STZ theory, compaction results from
elementary rearrangements, involving several molecules at the
mesoscopic scale'', rather than ``the motion of a single grain in a
hole''~\cite{lemaitre02b}.  Of course, we have reached the same
conclusion in the completely different context of granular flow.
Lema\^itre gives no further description of the ``elementary
rearrangements'' associated with free-volume dynamics, but the Spot
Model provides a possible mathematical framework, which also naturally
accounts for diffusion. This seems like an interesting direction for
future research.

\section{ Conclusion }
\label{sec:concl}

Returning to our original question about random walks, we arrive at a
simple {\it dynamical} distinction between different states of 
matter:
\begin{description}
\item[Gas.] The particles in a gas undergo independent random walks,
as ballistic trajectories are randomly redirected by collisions.
\item[Crystal.] The particles in a (single) crystal diffuse
by hopping to either interstitial or vacant lattice sites.
Both interstitials and vacancies undergo thermally activated,
independent random walks on the lattice (at sufficiently low defect
density). 
\item[Liquid.] The particles in a liquid move cooperatively 
at the scale of several particles, but, due to
the high internal kinetic energy, they undergo independent random
walks over much longer distances at experimental time scales.
\item[Amorphous.] The particles in an amorphous state move
cooperatively at experimental time scales, due to the low internal
energy. Proximity to jamming precludes the vacancy/interstitial
mechanism (inserting particles into holes). Instead, particles undergo
locally correlated random walks in response to diffusing spots of free
volume.
\end{description}
This is surely an oversimplification. For example, directional
covalent bonding can significantly alter the picture: In amorphous
silicon, vacancy (dangling bond~\cite{burlakov01}) or interstitial
(floating bond~\cite{pantelides86}) mechanisms may dominate
relaxation, while in open lattices like diamond silicon, cooperative
self-diffusion can also occur (concerted
exchange~\cite{pandey86}). Nevertheless, for materials with dense
disordered phases, the state classification above by diffusion
mechanism seems reasonable.

In summary, the Spot Model provides a simple mathematical framework
for cooperative diffusion, when independent random walks are inhibited
by packing constraints. Like the original random-walk concept, which
is now used in many fields other than physics, the basic idea of a
moving ``spot of influence'' may also have broader applications.

\vspace{0.2in}

\section*{Acknowledgements}

This work was supported by the U. S. Department of Energy (grant
DE-FG02-02ER25530) and the Norbert Wiener Research Fund and NEC Fund
at MIT.  The author is grateful to J. Choi, C. Gu\'aqueta,
A. Kudrolli, R. R. Rosales, C. H. Rycroft, and A. Samadani for
providing figures of unpublished results and for many stimulating
discussions, and to A. S. Argon, L. Bocquet, M. Demkowicz, R. Raghavan for
references to the glass literature.

\bibliography{spotmodel}

\end{document}